\documentclass[11pt, a4paper]{moonbit}

\usepackage[authoryear, sort&compress, round]{natbib}
\usepackage{dblfloatfix}
\usepackage{subcaption}
\usepackage{booktabs}
\usepackage{tabularx}
\usepackage{longtable}
\usepackage{multirow}
\usepackage{tikz}
\usepackage{pgfplots}
\usepackage{cleveref}
\usepackage{listings}
\usepackage{xspace}
\usepackage{placeins}
\usepackage{float}

\pgfplotsset{compat=1.18}
\usetikzlibrary{arrows.meta, positioning, shapes.geometric}

\providecommand{\absfont}{\linespread{1.2}\fontsize{11}{12}\selectfont}

\title{SWE-AGI:\@ Benchmarking Specification-Driven Software Construction with MoonBit in the Era of Autonomous Agents}

\author[1,\equalcontribmark]{Zhirui Zhang}
\author[1,2,\equalcontribmark]{Hongbo Zhang}
\author[1,\equalcontribmark]{Haoxiang Fei}
\author[1]{Zhiyuan Bao}
\author[1]{Yubin Chen}
\author[1]{Zhengyu Lei}
\author[1]{Ziyue Liu}
\author[1]{Yixuan Sun}
\author[1]{Mingkun Xiao}
\author[1]{Zihang Ye}
\author[1]{Yu Zhang}
\author[1]{Hongcheng Zhu}
\author[1]{Yuxiang Wen}
\author[1,2,\correspondingmark]{Heung-Yeung Shum}
\affil[1]{International Digital Economy Academy}
\affil[2]{The Hong Kong University of Science and Technology}
\affil[1]{\emails{zrustc11@gmail.com, hzhangcy@connect.ust.hk, feihaoxiang@idea.edu.cn}}
\affilnote{\equalcontribmark}{Equal contribution.}
\affilnote{\correspondingmark}{Corresponding author: \emails{hshum@ust.hk}}%

\date{}

\begin{document}

\begin{abstract}

Although large language models (LLMs) have demonstrated impressive coding capabilities, their ability to autonomously build production-scale software from explicit specifications remains an open question.
In this paper, we introduce SWE-AGI, the first open-source benchmark for evaluating end-to-end, specification-driven construction of software systems written in MoonBit.
SWE-AGI tasks require LLM-based agents to implement a range of software systems, including parsers, interpreters, binary decoders, and SAT solvers, strictly from authoritative standards and RFCs under a fixed API scaffold.
Each task involves implementing $10^3$–$10^4$ lines of core logic, corresponding to weeks or months of engineering effort for an experienced human developer. 
By leveraging the nascent MoonBit ecosystem, SWE-AGI minimizes data leakage, forcing agents to rely on long-horizon architectural reasoning rather than code retrieval.
Across frontier models, gpt-5.3-codex achieves the best overall performance (solving 19/22 tasks, 86.4\%), outperforming claude-opus-4.6 (15/22, 68.2\%), and kimi-2.5 exhibits the strongest performance among open-source models.
However, performance degrades sharply with increasing task difficulty, particularly on hard, specification-intensive systems.
Behavioral analysis further reveals that as codebases scale, code reading, rather than writing, becomes the dominant bottleneck in AI-assisted development.
Overall, while specification-driven autonomous software engineering is increasingly viable, substantial challenges remain before it can reliably support production-scale development.

\end{abstract}

\maketitle

\section{Introduction}

Large language models (LLMs) \citep{openai2025gpt5,google2025gemini25,anthropic2025claude4,deepseek2025v32,qwen2025qwen3,kimi2025k2} are increasingly deployed as software engineering (SWE) agents: they read specifications, write and refactor code, run tests, and iterate over long trajectories.
As this workflow becomes a practical interface for building and maintaining software, evaluation must move past single-shot code completion to address a more fundamental challenge: can an AI system autonomously carry out a production-scale implementation from explicit requirements to generate a correct, robust, and maintainable codebase?

Most existing benchmarks only partially capture this end-to-end capability.
Function- and problem-level tasks \citep{chen2021evaluating,austin2021program} are often short-horizon and can be solved via pattern matching or overfitting to limited tests.
Repository-issue benchmarks \citep{jimenez2023swebench,deng2025swebenchpro,yang2024sweagent} more closely reflect iterative development, but their results are frequently confounded by repository-specific conventions, hidden degrees of freedom in tooling, and difficult-to-control training-data overlap.
To measure autonomy at this level, a benchmark should instead be specification-grounded, production-scale, and evaluated using deterministic, human-validated tests under a standardized interface.

In this paper, we introduce SWE-AGI\footnote{\url{https://github.com/moonbitlang/SWE-AGI}}, the first open-source benchmark for assessing autonomous software engineering through specification-driven, from-scratch system construction in MoonBit, a modern programming language with a nascent ecosystem.
Leveraging MoonBit's native support for spec-first development and its integrated toolchain \citep{moonbit}, SWE-AGI tasks require LLM-based agents to implement production-grade, standards-compliant systems in MoonBit strictly from authoritative specifications within a fixed API scaffold.
Concretely, MoonBit supports declaration-first workflows via the \texttt{declare} keyword, which allows developers to write function signatures and type declarations first and provide implementations later.
Combined with the unified build/test/package workflow (\texttt{moon}), this yields a standardized end-to-end engineering workflow that closely matches real-world practice.
Since SWE-AGI focuses on production-scale software systems that are largely absent from the current MoonBit ecosystem (e.g., a CDCL SAT solver, a WASM decoder/validator, and a standards-compliant C99 parser), it explicitly prioritizes \emph{reasoning over retrieval}: success depends on sustained specification understanding, architectural decision-making, and disciplined long-horizon implementation rather than recalling near-matching reference code.

SWE-AGI targets production-scale software engineering and consists of 22 tasks spanning seven categories.
These tasks are stratified into three difficulty tiers based on code volume and implementation complexity, comprising 6 easy, 8 medium, and 8 hard tasks.
Completing the core logic of a SWE-AGI task requires $10^3$--$10^4$ lines of implementation under a fixed API scaffold, corresponding to weeks to months of engineering effort for an experienced human developer.
To support evaluation at this scale, each task provides normative specifications (\texttt{specs/}), an explicit task statement (\texttt{TASK.md}), and a visible public test subset for local iteration, while benchmark scoring is performed solely on final submissions evaluated against hidden private tests.
This evaluation design shifts the challenge from isolated code generation to an end-to-end software engineering process, requiring agents to demonstrate sustained autonomy rather than relying on one-shot generation: interpreting complex specifications, becoming familiar with MoonBit, architecting modular systems, and performing self-directed testing.

In our latest evaluation, gpt-5.3-codex achieves the strongest overall performance (solving 19/22 tasks, 86.4\%), outperforming gpt-5.2-codex (17/22, 77.3\%), claude-opus-4.6 (15/22, 68.2\%), and claude-opus-4.5 (10/22, 45.5\%).
Although these frontier agents successfully complete all easy-tier tasks, performance degrades on the medium and hard tiers as task difficulty increases: success rates for both gpt-5.3-codex and gpt-5.2-codex decline sharply on hard tasks, whereas claude-opus-4.6 and claude-opus-4.5 begin to falter from the medium tier onward.
In addition, we evaluate several other LLMs on six easy-tier tasks, including gemini-3-flash, kimi-k2.5, claude-sonnet-4.5, deepseek-v3.2, glm-4.7, and qwen3-max.
Most of these models solve at most 2/6 easy tasks, revealing a substantial performance gap relative to the evaluated frontier agents even at the lowest difficulty level.
Among these easy-tier baselines, kimi-k2.5 achieves the highest overall test-suite pass rate (92.0\%) while tying for the best task success rate (2/6).
We further conduct a behavioral analysis of end-to-end SWE agents and observe that code reading, rather than code writing, emerges as the central bottleneck in AI-assisted software development.
As codebases scale, maintaining a coherent modular architecture becomes the dominant activity.
Consistent with this observation, gpt-5.2-codex allocates a larger fraction of its actions to code understanding, while gpt-5.3-codex exhibits a more iteration-oriented profile with higher debugging share and substantially fewer logged actions, improving time-to-solution and overall task completion efficiency.
Overall, these results suggest that autonomous software engineering from explicit specifications is becoming increasingly feasible, yet remains far from a solved problem at production scale.

This paper makes three contributions:
\begin{itemize}
  \item We introduce SWE-AGI, the first benchmark focusing on the end-to-end construction of complex systems from authoritative standards. It shifts the evaluation paradigm from localized code completion to long-horizon architectural reasoning and rigorous system implementation.
  \item We design a specification-grounded, retrieval-resistant evaluation setting by leveraging MoonBit's nascent ecosystem and spec-first primitives, ensuring that success reflects genuine long-horizon engineering capabilities rather than recall of near-matching artifacts.
  \item We benchmark state-of-the-art SWE agents built on frontier LLMs on SWE-AGI and present a comprehensive empirical and behavioral analysis, revealing strong performance on easy-tier tasks but substantial degradation as task difficulty increases.
\end{itemize}

\begin{figure}[t]
\vspace*{-2mm}
\centering
\includegraphics[width=\linewidth]{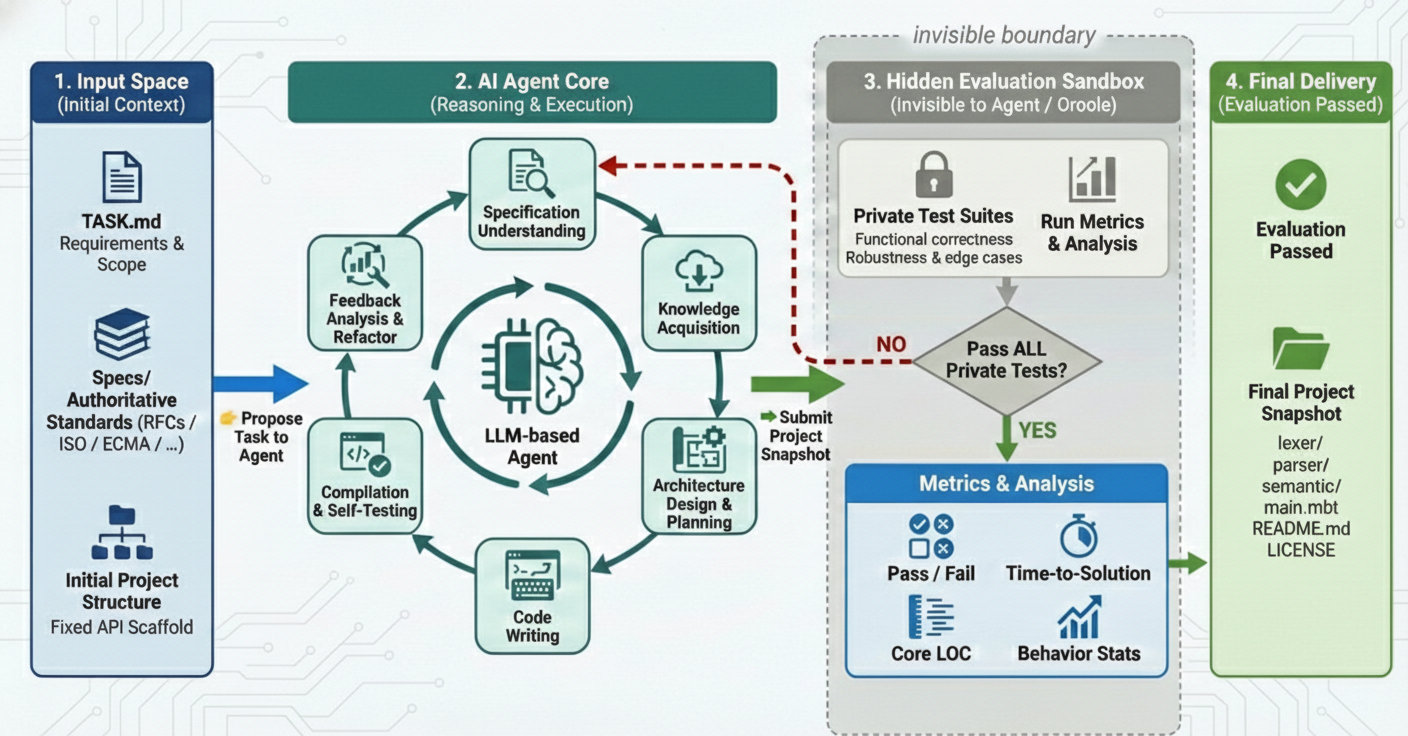}
\caption{SWE-AGI benchmark execution pipeline. From a cold-start starter repository (inputs: \texttt{TASK.md}, normative \texttt{specs/}, a MoonBit scaffold, and public tests), an autonomous agent iterates over design/implementation and local testing, submits the project for evaluation (via \texttt{swe-agi-submit}), receives pass/fail feedback, and repeats until a verified submission passes.}
\label{fig:execution-pipeline}
\end{figure}

\section{SWE-AGI Benchmark}
SWE-AGI evaluates autonomous software engineering through \emph{specification-driven, from-scratch} construction of production-scale systems under a fixed MoonBit scaffold.
Section~\ref{sec:task-formulation} defines the per-task interface and agent execution loop, while Section~\ref{sec:construction} describes the benchmark construction process.

\subsection{Task Formulation}
\label{sec:task-formulation}

Figure~\ref{fig:execution-pipeline} illustrates the SWE-AGI execution pipeline.
Each task is framed as the construction of a complete software system \emph{from explicit specifications} (e.g., RFCs and standards) under a fixed MoonBit API scaffold.
Concretely, a task is distributed as a starter repository that provides: (i) an explicit task statement (\texttt{TASK.md}) with acceptance criteria, constraints, and executable instructions; (ii) normative references (\texttt{specs/}); (iii) declaration-first API scaffolding that fixes the public interface; and (iv) a visible public test subset for fast local iteration.
These components collectively define the core loop of the AI agent: interpreting the specifications, implementing against a fixed interface, validating locally, and iteratively submitting until the hidden private tests pass.

Evaluation considers only final submissions against hidden private tests, allowing agents full freedom in intermediate reasoning, testing, and implementation strategies.
Private tests reduce overfitting to the visible suite and enforce specification-grounded implementations, while preserving an iterative, real-world-like engineering loop.
During development, agents may supplement the provided public tests with their own spec-grounded checks, perform local validation via \texttt{moon test}, and iteratively submit solutions using \texttt{swe-agi-submit} until the submission passes the private test suite.

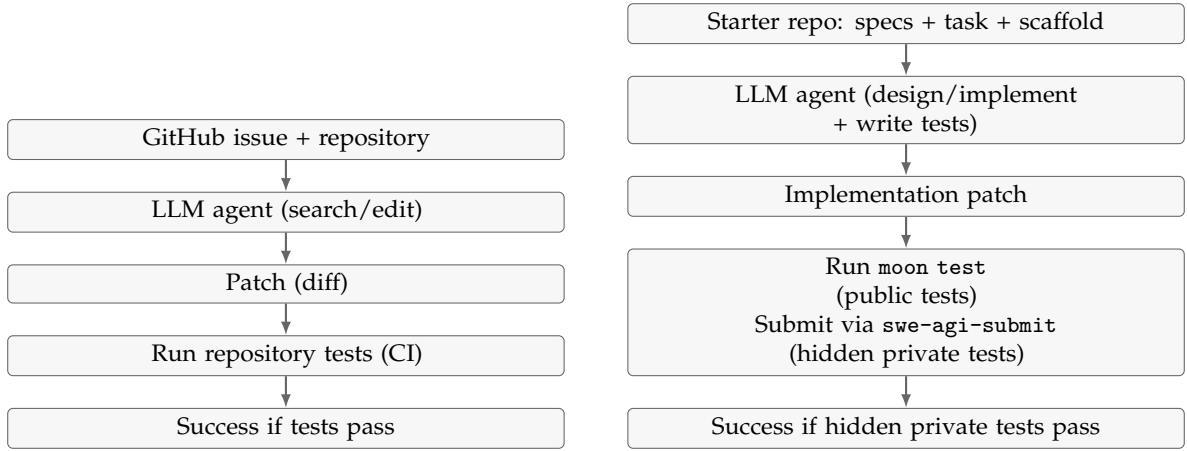
\begin{figure}[t]
  \vspace*{-2mm}
  \centering
  \begin{subfigure}[t]{0.49\linewidth}
    \centering
    \begin{tikzpicture}[
        node distance=4.2mm,
        box/.style={draw=black!60, fill=black!3, rounded corners=2pt, align=center, inner xsep=6pt, inner ysep=3pt, font=\footnotesize, text width=0.88\linewidth},
        arrow/.style={-{Latex[length=1.8mm]}, thick, draw=black!60}
      ]
      \node[box] (issue) {GitHub issue + repository};
      \node[box, below=of issue] (agent) {LLM agent (search/edit)};
      \node[box, below=of agent] (patch) {Patch (diff)};
      \node[box, below=of patch] (tests) {Run repository tests (CI)};
      \node[box, below=of tests] (score) {Success if tests pass};

      \draw[arrow] (issue) -- (agent);
      \draw[arrow] (agent) -- (patch);
      \draw[arrow] (patch) -- (tests);
      \draw[arrow] (tests) -- (score);
    \end{tikzpicture}
    \caption{SWE-bench-style: issue resolution in existing repositories.}
  \end{subfigure}
  \hfill
  \begin{subfigure}[t]{0.49\linewidth}
    \centering
    \begin{tikzpicture}[
        node distance=4.2mm,
        box/.style={draw=black!60, fill=black!3, rounded corners=2pt, align=center, inner xsep=6pt, inner ysep=3pt, font=\footnotesize, text width=0.88\linewidth},
        arrow/.style={-{Latex[length=1.8mm]}, thick, draw=black!60}
      ]
      \node[box] (input) {Starter repo: specs + task + scaffold};
      \node[box, below=of input] (agent) {LLM agent (design/implement\\+ write tests)};
      \node[box, below=of agent] (patch) {Implementation patch};
      \node[box, below=of patch] (tests) {Run {\scriptsize\texttt{moon test}}\\(public tests)\\Submit via {\scriptsize\texttt{swe\allowbreak-agi\allowbreak-submit}}\\(hidden private tests)};
      \node[box, below=of tests] (score) {Success if hidden private tests pass};

      \draw[arrow] (input) -- (agent);
      \draw[arrow] (agent) -- (patch);
      \draw[arrow] (patch) -- (tests);
      \draw[arrow] (tests) -- (score);
    \end{tikzpicture}
    \caption{SWE-AGI style: specification-driven implementation (with agent-written tests) in a fixed scaffold.}
  \end{subfigure}
  \caption{Conceptual contrast between SWE-bench \citep{jimenez2023swebench} and SWE-AGI evaluation settings.}
  \label{fig:swebench-vs-sweagi}
\end{figure}

\begin{table}[t]
  \centering
  \caption{Comparison of SWE-AGI to representative coding and software engineering benchmarks (high-level characterization; code scale and workload are rough order-of-magnitude indicators).}
  \label{tab:comparison}
  \scriptsize
  \begin{tabularx}{0.97\linewidth}{@{}>{\centering\arraybackslash}p{2.15cm} >{\centering\arraybackslash}X >{\centering\arraybackslash}X >{\centering\arraybackslash}X >{\centering\arraybackslash}X >{\centering\arraybackslash}X@{}}
    \toprule
    Benchmark & Primary goal & Typical code scale & Workload & Difficulty focus & Evaluation criteria \\
    \midrule
    HumanEval\\\citep{chen2021evaluating} & Function synthesis & $\sim 10^1$ LOC & minutes & Local correctness & Unit tests \\
    \midrule[0.2pt]
    MBPP\\\citep{austin2021program} & Small programs & $\sim 10^1$--$10^2$ LOC & minutes--hours & Edge cases; basic reasoning & Unit tests \\
    \midrule[0.2pt]
    APPS\\\citep{hendrycks2021apps} & Programming problems & $\sim 10^2$--$10^3$ LOC & hours & Problem solving; I/O behavior & Test-based \\
    \midrule[0.2pt]
    LiveCodeBench\\\citep{jain2024livecodebench} & Programming problems (time-based) & $\sim 10^2$--$10^3$ LOC & hours & Contamination-resistant coding skill & Test-based; time-evolving set \\
    \midrule[0.2pt]
    RepoBench\\\citep{liu2023repobench} & Repository-level completion & $\sim 10^1$--$10^2$ LOC & seconds--minutes & Cross-file context retrieval & Completion accuracy \\
    \midrule[0.2pt]
    SWE-bench\\\citep{jimenez2023swebench} & Repo issue resolution & $\sim 10^1$--$10^3$ LOC & hours--days & Debugging; tool use; integration & Repository tests (CI) \\
    \midrule[0.2pt]
    SWE-bench Pro\\\citep{deng2025swebenchpro} & Repo issue resolution (enhanced) & $\sim 10^1$--$10^3$ LOC & hours--days & Debugging; improved coverage & Repository tests (CI) \\
    \midrule[0.2pt]
    SWE-AGI & Autonomous SWE from explicit specifications & $\sim 10^3$--$10^4$ LOC & weeks--months & Spec comprehension; system design & Hidden private tests via submission\\
    \bottomrule
  \end{tabularx}
\end{table}

Figure~\ref{fig:swebench-vs-sweagi} contrasts SWE-AGI with SWE-bench--style issue resolution in existing repositories.
Compared to common coding benchmarks in Table~\ref{tab:comparison}, SWE-AGI shifts the primary sources of difficulty toward specification reading and operationalization, long-horizon system design and multi-module implementation, and iterative debugging/refactoring under build/test feedback in an open development setting\footnote{External tools such as web search may be used, but are less helpful when near-matching implementations are unavailable.}.
Each task typically requires implementing $10^3$--$10^4$ lines of core logic, corresponding to weeks or months of engineering effort for an experienced human developer, and is accompanied by high-coverage, human-validated test suites that evaluate both functional correctness on well-formed inputs and robustness to malformed inputs.

\subsection{Benchmark Construction}
\label{sec:construction}
SWE-AGI consists of 22 tasks spanning seven categories: (i) Template and Domain-Specific Languages (pug, jq); (ii) Data Serialization and Configuration Formats (csv, ini, yaml, toml); (iii) Markup and Document Formats (xml, html5); (iv) Programming Language Front-Ends (c99, lua, ecma262, python, r6rs); (v) Binary Formats and Streaming Decoders (git\_object, protobuf, zip, capnp, wasm); (vi) Networking and Protocol State Machines (uri, hpack, url); and (vii) Automated Reasoning and SAT Solving (cdcl).
Each task is framed as an end-to-end software system with a fixed API scaffold.
Tasks are assigned to three coarse difficulty tiers (\emph{Easy}/\emph{Medium}/\emph{Hard}), primarily based on the estimated scale of core implementation code (excluding tests), and further informed by semantic complexity indicators such as multi-phase parsing and validation, large state machines, and strict error-recovery requirements.
Appendix~\ref{app:tasks} provides detailed task descriptions, per-task difficulty assignments, and overall tier counts.

SWE-AGI prioritizes \emph{reasoning over retrieval} and is explicitly designed to minimize superficial success through memorization or direct code reuse.
Accordingly, we focus on systems that are largely absent from the current MoonBit ecosystem and that demand sustained engagement with formal specifications and non-trivial engineering decisions, including interface design, data-structure selection, and robust error handling.

\paragraph{Repository packaging.}
Following the interface defined in Section~\ref{sec:task-formulation}, tasks are constructed by selecting authoritative upstream specifications (e.g., standards, RFCs, and reference documents), distilling explicit acceptance criteria---including corner cases and error semantics---into \texttt{TASK.md}, and providing a fixed API scaffold together with high-coverage test suites.
The test suites comprise a visible public subset for local iteration and a hidden private subset for verification.
To support both agent usability and researcher auditability, each task directory includes normative references (\texttt{specs/}), a single task entry point (\texttt{TASK.md}), a minimal MoonBit package configuration (\texttt{moon.mod.json} and \texttt{moon.pkg.json}), and scaffolded declarations (typically in \texttt{*\_spec.mbt}) that define and freeze the public API\@.
Overall, tasks are packaged to minimize hidden requirements and evaluation variance, ensuring that success depends on specification-grounded engineering rather than repository-specific conventions.
A typical directory layout is shown in Listing~\ref{lst:task-layout}.

\begin{lstlisting}[float, floatplacement=tbp, caption={Typical directory layout for a SWE-AGI task.}, label={lst:task-layout}]
tasks/<task>/
  specs/                  # upstream specs and reference documents
  TASK.md                 # goal, scope, API, behavioral rules, test execution
  *_spec.mbt              # fixed API declarations + helper contracts
  *_pub_test.mbt          # public tests (subset of full suite)
  *_priv_test.mbt         # private tests (held out; only in evaluation checkout)
  moon.mod.json           # package manifest and dependencies
  moon.pkg.json           # package lockfile (pinned deps)
\end{lstlisting}

\paragraph{Test sets and evaluation metrics.}
Tests in SWE-AGI are constructed through a hybrid process.
Canonical cases are adapted from authoritative specifications and reference materials, and are expanded with systematic edge cases---including property-based generators, LLM-generated candidates, and fuzz-style mutations where appropriate---followed by manual triage to ensure specification-consistent expectations.
SWE-AGI reports both \emph{functional} and \emph{engineering} metrics (Table~\ref{tab:metrics}).
Functional performance is measured by task success rate and test-suite pass rate (overall), while engineering effort and efficiency are characterized by time to solution and implementation size (core LOC), respectively.
In addition, we report behavioral statistics to support more detailed analysis of agent behavior.
Performance metrics such as runtime and memory usage are not scored in the current release, but are reserved for future versions once state-of-the-art models achieve consistently high task success rates.

\begin{table}[t]
  \centering
  \small
  \caption{Recommended SWE-AGI metrics for reporting.}
  \label{tab:metrics}
  \begin{tabularx}{\linewidth}{@{}lX@{}}
    \toprule
    Metric & Definition \\
    \midrule
    Task success rate & Fraction of tasks for which the final submission compiles successfully and passes all hidden private tests under the specified evaluation protocol. \\
    Test-suite pass rate (overall) & Pass rate on the full evaluation test suite executed by the evaluator (public+private), reported as passed/total with no public/private split. \\
    Time to solution & Wall-clock time to the first submission that passes the hidden private tests. \\
    Implementation size (core LOC) & Number of non-test MoonBit lines of code, excluding public and private tests as well as auxiliary tooling, used as a coarse proxy for system scale. \\
    Behavior stats (optional) & Aggregated distribution of agent actions over the engineering loop (e.g., specification reading, code reading and writing, debugging, test execution, planning or navigation, and external search), computed from tool usage logs. \\
    \bottomrule
  \end{tabularx}
\end{table}

\subsection{Language Choice: MoonBit}
\label{sec:features}
SWE-AGI adopts MoonBit \citep{moonbit} as its implementation language to control distributional bias during evaluation.
As a relatively new programming language with a still-maturing ecosystem, MoonBit is largely absent from existing large-scale pretraining corpora and public code repositories.
This reduces the likelihood that agents can exploit memorized near-solutions or ecosystem-specific shortcuts, thereby shifting the evaluation signal toward specification comprehension, algorithmic reasoning, and correct end-to-end implementation.

MoonBit's type soundness and unified toolchain further improve the quality and timeliness of feedback available to autonomous agents.
Its emphasis on data-oriented programming, immutability, and exhaustive pattern matching surfaces many classes of errors---such as missing cases, violated invariants, and type mismatches---at compile time rather than at runtime.
Moreover, MoonBit implementations are often more concise for a given specification, reducing overall code volume and the surface area for latent bugs.
Combined with fast compilation\footnote{In reported benchmarks, MoonBit can compile hundreds of packages in approximately one second, substantially reducing iteration overhead compared to traditional programming languages.} and test execution via the \texttt{moon} toolchain, these properties enable high-frequency compile--test--refine cycles with low feedback latency, providing earlier and more actionable signals within the agent loop.

Finally, MoonBit's built-in support for separating interface and implementation enables a scaffolded evaluation setup in which public APIs, type signatures, and module boundaries are explicitly fixed using \texttt{declare} (Figure~\ref{fig:declare-code}).
Agents are required to implement the specified interfaces exactly, with deviations detected at compile time rather than implicitly tolerated at runtime.
This enforces clear boundaries, prevents interface-level circumvention, and ensures that evaluation focuses on the correctness and robustness of the implemented logic rather than flexibility in interface design.

\begin{figure}
  \begin{minipage}[t]{\textwidth}
\begin{lstlisting}[basicstyle=\ttfamily\small,frame=single,backgroundcolor=\color{gray!10}]
declare pub(all) type CProgram
/// Parse a C99 translation unit from source text.
declare pub fn parse(code : StringView) -> CProgram raise
/// Encode the parsed program into the explicit test JSON schema
declare pub fn CProgram::to_test_json(self : CProgram) -> Json
\end{lstlisting}
  \end{minipage}
  \caption{Declaration-first, spec-driven workflow in MoonBit. The {\small\texttt{declare}} keyword fixes public types and function signatures (e.g., parser entry points and test-schema encoders) before implementation.}
  \label{fig:declare-code}
\end{figure}

\section{Evaluation of Frontier Agents}
\label{sec:eval}

We evaluate software engineering agents built on frontier models on SWE-AGI under an open development setting in which the scored private tests are hidden from the model.
Throughout, we use \emph{model} to refer to the underlying LLM, and \emph{agent} to refer to the model coupled with an execution front-end, tool access, and associated policies.
Agents must translate \texttt{TASK.md} plus authoritative references (\texttt{specs/}) into a working MoonBit implementation under a fixed scaffold, iterate locally using public tests (10\% of all tests), and submit via \texttt{swe-agi-submit} until the evaluator reports that hidden private tests pass.

\subsection{Setup}
\label{sec:eval-setup}
\begin{sloppypar}
We evaluate each model via an agent front-end that can edit the repository, execute local commands, and iteratively submit solutions.
We use Codex CLI with gpt-5.3-codex and gpt-5.2-codex\footnote{For Codex CLI, we run gpt-5.3-codex in \textit{xhigh} thinking mode. For gpt-5.2-codex, we adopt \textit{high} thinking mode, since \textit{xhigh} incurred prohibitively long wall-clock runtimes.}; Gemini CLI with gemini-3-flash\footnote{In Gemini CLI runs, we observe repeated execution failures, including three instances of ``Loop detected, stopping execution'' and two instances of ``[API Error: Premature close]'', which resulted in a low task pass rate. Due to these stability issues, we omit results for gemini-3-pro from our reported evaluations.}; Claude Code with claude-opus-4.6, claude-opus-4.5, claude-sonnet-4.5, qwen3-max\footnote{qwen3-max-thinking (2026-01-23)}, glm-4.7, and deepseek-v3.2\footnote{deepseek-reasoner}; and Kimi CLI with kimi-k2.5\footnote{kimi-k2.5-thinking}.
We will release the execution scripts\footnote{\url{https://github.com/moonbitlang/SWE-AGI}} along with the model outputs and corresponding run logs\footnote{\url{https://github.com/moonbitlang/SWE-AGI-Eval}} to support reproducibility.
\end{sloppypar}

A task is considered \emph{passed} if the final submitted project compiles and the evaluator reports zero failed hidden private tests in a clean checkout; otherwise it is \emph{failed}. In addition to task-level success, we report test-suite pass rates (overall), wall-clock duration to the final submission (hours), implementation size (core LOC, excluding tests), and token usage aggregated from tool logs.
We conduct full evaluations for gpt-5.3-codex, gpt-5.2-codex, claude-opus-4.6, and claude-opus-4.5.
In addition, we conduct a rapid assessment of agentic coding capabilities on six easy-tier tasks for claude-sonnet-4.5, kimi-k2.5, glm-4.7, gemini-3-flash, deepseek-v3.2, and qwen3-max.
Given the low easy-tier success rates, we limit these additional evaluations to the easy tier and do not extend testing to higher difficulty levels. We do not enforce an explicit budget constraint; instead, we report token consumption and wall-clock time as post hoc efficiency metrics aggregated per run from the recorded tool logs. For Claude Code executions (claude-opus-4.6 and claude-opus-4.5), we additionally report per-task monetary costs extracted from the agent logs in the detailed per-task tables.

\begin{table}[!htbp]
  \centering
  \caption{Evaluation summary by difficulty tier.}
  \label{tab:eval-summary}
  \scriptsize
  \setlength{\tabcolsep}{3pt}
  \resizebox{\linewidth}{!}{%
    \begin{tabular}{@{}llcrrr@{}}
      \toprule
      Difficulty & Model & Tasks Passed & Test-suite Pass Rate & Avg.\ Time & Avg.\ Core LOC\\
      \midrule
      \multirow[c]{10}{*}{Easy} & gpt-5.3-codex & \textbf{6/6} & \textbf{100.0\% (avg of 6; total 1604/1604)} & 0.28h & 1305\\
      & gpt-5.2-codex & \textbf{6/6} & \textbf{100.0\% (avg of 6; total 1604/1604)} & 0.81h & 1081\\
      & claude-opus-4.6 & \textbf{6/6} & \textbf{100.0\% (avg of 6; total 1604/1604)} & 0.45h & 1781\\
      & claude-opus-4.5 & \textbf{6/6} & \textbf{100.0\% (avg of 6; total 1604/1604)} & 0.39h & 1092\\
      & claude-sonnet-4.5 & 0/6 & 76.1\% (avg of 6; total 556/1604) & 0.32h & 930\\
      & kimi-k2.5 & 2/6 & 92.0\% (avg of 6; total 1338/1604) & 0.99h & 1163\\
      & glm-4.7 & 2/6 & 64.2\% (avg of 6; total 456/1604) & 0.70h & 904\\
      & gemini-3-flash & 2/6 & 49.8\% (avg of 6; total 376/1604) & 0.25h & 558\\
      & deepseek-v3.2 & 1/6 & 16.7\% (avg of 6; total 138/1604) & 3.4h & 1070\\
      & qwen3-max & 0/6 & 13.9\% (avg of 6; total 94/1604) & 2.6h & 850\\
      \midrule
      \multirow[c]{4}{*}{Medium} & gpt-5.3-codex & \textbf{8/8} & \textbf{100.0\% (avg of 8; total 5284/5284)} & 1.2h & 2575\\
      & gpt-5.2-codex & \textbf{7/8} & \textbf{98.9\% (avg of 8; total 5176/5284)} & 5.1h & 4702\\
      & claude-opus-4.6 & 5/8 & 93.6\% (avg of 8; total 5146/5284) & 3.5h & 4867\\
      & claude-opus-4.5 & 3/8 & 82.6\% (avg of 8; total 4593/5284) & 1.3h & 3304\\
      \midrule
      \multirow[c]{4}{*}{Hard} & gpt-5.3-codex & \textbf{5/8} & \textbf{87.9\% (avg of 8; total 13976/15638)} & 1.7h & 6255\\
      & gpt-5.2-codex & \textbf{4/8} & \textbf{91.2\% (avg of 8; total 13205/15638)} & 7.8h & 9034\\
      & claude-opus-4.6 & 4/8 & 81.1\% (avg of 8; total 14625/15638) & 5.7h & 10103\\
      & claude-opus-4.5 & 1/8 & 67.0\% (avg of 8; total 10621/15638) & 1.7h & 6603\\
      \bottomrule
    \end{tabular}%
  }
\end{table}

\begingroup
\renewcommand{\topfraction}{0.95}
\renewcommand{\textfraction}{0.05}
\renewcommand{\floatpagefraction}{0.85}
\setlength{\textfloatsep}{6pt plus 2pt minus 2pt}
\renewcommand{\arraystretch}{0.95}
\begin{table}[!htbp]
  \centering
  \caption{Per-task detailed results for gpt-5.3-codex and gpt-5.2-codex. Tokens report input/output tokens as logged; for Codex CLI we report \textit{input\_tokens} (excluding \textit{cached\_input\_tokens}). Cost reports per-task dollar cost; values for Codex CLI are approximate (using API price), and this estimate is inaccurate since it ignores the overhead introduced by reasoning tokens. Due to the excessively long runtime of \textit{ecma262} (exceeding 42 hours), we evaluate it using a 42-hour snapshot. As the execution did not finish within this window, input/output token statistics are unavailable in the logs and are reported as \textit{N/A}.}
  \label{tab:eval-detail-gpt}
  \footnotesize
  \setlength{\tabcolsep}{3pt}
  \resizebox{0.93\linewidth}{!}{%
    \begin{tabular}{@{}l l c r r r r r@{}}
      \toprule
      Task & Model & Task Passed & Test-suite Pass Rate & Duration & Core LOC & Tokens (in/out) & Cost (\$)\\
      \midrule
      \multirow[c]{2}{*}{\texttt{pug}} & gpt-5.3-codex & Yes & \textbf{100.0\% (251/251)} & 3.5h & 2709 & 92.73M/319.3k & \(\sim\)22.69\\
      & gpt-5.2-codex & Yes & \textbf{100.0\% (251/251)} & 24.6h & 14251 & 647.72M/3.28M & \(\sim\)176.90\\
      \midrule
      \multirow[c]{2}{*}{\texttt{jq}} & gpt-5.3-codex & Yes & \textbf{100.0\% (218/218)} & 1.1h & 4914 & 45.68M/184.9k & \(\sim\)11.41\\
      & gpt-5.2-codex & Yes & \textbf{100.0\% (218/218)} & 1.9h & 6416 & 68.93M/267.2k & \(\sim\)16.67\\
      \midrule
      \multirow[c]{2}{*}{\texttt{csv}} & gpt-5.3-codex & Yes & \textbf{100.0\% (98/98)} & 0.21h & 467 & 5.79M/37.1k & \(\sim\)1.85\\
      & gpt-5.2-codex & Yes & \textbf{100.0\% (98/98)} & 0.68h & 440 & 12.97M/78.9k & \(\sim\)3.69\\
      \midrule
      \multirow[c]{2}{*}{\texttt{ini}} & gpt-5.3-codex & Yes & \textbf{100.0\% (98/98)} & 0.26h & 1495 & 9.16M/46.9k & \(\sim\)2.57\\
      & gpt-5.2-codex & Yes & \textbf{100.0\% (98/98)} & 0.77h & 927 & 17.29M/104.7k & \(\sim\)4.95\\
      \midrule
      \multirow[c]{2}{*}{\texttt{yaml}} & gpt-5.3-codex & Yes & \textbf{100.0\% (345/345)} & 0.90h & 856 & 26.04M/148.3k & \(\sim\)47.64\\
      & gpt-5.2-codex & Yes & \textbf{100.0\% (345/345)} & 3.9h & 3664 & 95.40M/571.6k & \(\sim\)26.78\\
      \midrule
      \multirow[c]{2}{*}{\texttt{toml}} & gpt-5.3-codex & Yes & \textbf{100.0\% (733/733)} & 0.88h & 2149 & 20.80M/153.2k & \(\sim\)6.64\\
      & gpt-5.2-codex & Yes & \textbf{100.0\% (733/733)} & 3.0h & 2280 & 64.37M/345.0k & \(\sim\)17.25\\
      \midrule
      \multirow[c]{2}{*}{\texttt{xml}} & gpt-5.3-codex & Yes & \textbf{100.0\% (735/735)} & 1.2h & 4504 & 40.93M/165.3k & \(\sim\)10.57\\
      & gpt-5.2-codex & Yes & \textbf{100.0\% (735/735)} & 1.9h & 4946 & 61.76M/230.3k & \(\sim\)14.81\\
      \midrule
      \multirow[c]{2}{*}{\texttt{html5}} & gpt-5.3-codex & No & \textbf{86.2\% (7086/8221)} & 3.3h & 11080 & 67.40M/155.5k & \(\sim\)15.45\\
      & gpt-5.2-codex & No & 78.4\% (6444/8221) & 3.0h & 6433 & 51.76M/209.1k & \(\sim\)12.66\\
      \midrule
      \multirow[c]{2}{*}{\texttt{c99}} & gpt-5.3-codex & Yes & \textbf{100.0\% (117/117)} & 0.62h & 3624 & 22.06M/99.3k & \(\sim\)5.82\\
      & gpt-5.2-codex & Yes & \textbf{100.0\% (117/117)} & 3.1h & 5052 & 66.09M/332.6k & \(\sim\)17.68\\
      \midrule
      \multirow[c]{2}{*}{\texttt{lua}} & gpt-5.3-codex & Yes & \textbf{100.0\% (137/137)} & 0.93h & 9625 & 22.97M/163.2k & \(\sim\)7.09\\
      & gpt-5.2-codex & Yes & \textbf{100.0\% (137/137)} & 2.5h & 5574 & 71.14M/307.1k & \(\sim\)18.40\\
      \midrule
      \multirow[c]{2}{*}{\texttt{ecma262}} & gpt-5.3-codex & No & 17.5\% (108/618) & 0.75h & 7445 & 17.92M/109.6k & \(\sim\)5.45\\
      & gpt-5.2-codex & No & \textbf{97.7\% (604/618)} & 42.2h & 33302 & N/A & N/A\\
      \midrule
      \multirow[c]{2}{*}{\texttt{python}} & gpt-5.3-codex & Yes & \textbf{100.0\% (653/653)} & 2.3h & 7953 & 51.54M/228.4k & \(\sim\)13.86\\
      & gpt-5.2-codex & Yes & \textbf{100.0\% (653/653)} & 1.8h & 7675 & 64.68M/236.8k & \(\sim\)17.70\\
      \midrule
      \multirow[c]{2}{*}{\texttt{r6rs}} & gpt-5.3-codex & Yes & \textbf{100.0\% (1362/1362)} & 2.6h & 3751 & 52.46M/225.0k & \(\sim\)13.72\\
      & gpt-5.2-codex & No & 53.4\% (727/1362) & 2.9h & 6436 & 78.02M/349.6k & \(\sim\)20.19\\
      \midrule
      \multirow[c]{2}{*}{\texttt{git\_object}} & gpt-5.3-codex & Yes & \textbf{100.0\% (1000/1000)} & 0.44h & 840 & 9.85M/42.4k & \(\sim\)2.80\\
      & gpt-5.2-codex & Yes & \textbf{100.0\% (1000/1000)} & 1.2h & 1164 & 37.96M/148.5k & \(\sim\)9.78\\
      \midrule
      \multirow[c]{2}{*}{\texttt{protobuf}} & gpt-5.3-codex & Yes & \textbf{100.0\% (141/141)} & 0.19h & 1590 & 7.09M/34.4k & \(\sim\)2.05\\
      & gpt-5.2-codex & Yes & \textbf{100.0\% (141/141)} & 0.50h & 1670 & 14.26M/74.5k & \(\sim\)3.83\\
      \midrule
      \multirow[c]{2}{*}{\texttt{zip}} & gpt-5.3-codex & Yes & \textbf{100.0\% (1089/1089)} & 1.2h & 1258 & 41.19M/177.0k & \(\sim\)10.83\\
      & gpt-5.2-codex & Yes & \textbf{100.0\% (1089/1089)} & 3.2h & 1346 & 77.73M/373.3k & \(\sim\)20.32\\
      \midrule
      \multirow[c]{2}{*}{\texttt{capnp}} & gpt-5.3-codex & Yes & \textbf{100.0\% (111/111)} & 0.53h & 3114 & 17.83M/82.4k & \(\sim\)4.88\\
      & gpt-5.2-codex & Yes & \textbf{100.0\% (111/111)} & 1.2h & 2798 & 25.16M/136.3k & \(\sim\)7.04\\
      \midrule
      \multirow[c]{2}{*}{\texttt{wasm}} & gpt-5.3-codex & Yes & \textbf{100.0\% (800/800)} & 0.76h & 5085 & 29.51M/125.1k & \(\sim\)8.01\\
      & gpt-5.2-codex & Yes & \textbf{100.0\% (800/800)} & 2.1h & 3479 & 59.27M/270.4k & \(\sim\)15.67\\
      \midrule
      \multirow[c]{2}{*}{\texttt{uri}} & gpt-5.3-codex & Yes & \textbf{100.0\% (138/138)} & 0.25h & 1645 & 8.98M/45.2k & \(\sim\)2.63\\
      & gpt-5.2-codex & Yes & \textbf{100.0\% (138/138)} & 0.72h & 1128 & 15.90M/101.3k & \(\sim\)4.59\\
      \midrule
      \multirow[c]{2}{*}{\texttt{hpack}} & gpt-5.3-codex & Yes & \textbf{100.0\% (129/129)} & 0.33h & 1793 & 11.71M/56.7k & \(\sim\)3.10\\
      & gpt-5.2-codex & Yes & \textbf{100.0\% (129/129)} & 1.0h & 1157 & 23.90M/143.9k & \(\sim\)6.68\\
      \midrule
      \multirow[c]{2}{*}{\texttt{url}} & gpt-5.3-codex & Yes & \textbf{100.0\% (1220/1220)} & 0.32h & 926 & 11.36M/48.5k & \(\sim\)2.88\\
      & gpt-5.2-codex & No & 91.1\% (1112/1220) & 1.2h & 4849 & 31.86M/159.2k & \(\sim\)8.40\\
      \midrule
      \multirow[c]{2}{*}{\texttt{cdcl}} & gpt-5.3-codex & No & 99.6\% (4295/4312) & 2.2h & 1650 & 53.07M/96.7k & \(\sim\)11.14\\
      & gpt-5.2-codex & No & \textbf{99.8\% (4305/4312)} & 5.1h & 1380 & 47.34M/181.8k & \(\sim\)11.75\\
      \bottomrule
    \end{tabular}
  }
\end{table}

\begin{table}[!htbp]
  \centering
  \caption{Per-task detailed results for claude-opus-4.6 and claude-opus-4.5. Token and cost values are extracted from Claude Code logs when available; \textit{N/A} indicates missing token/cost logs (e.g., due to a Claude Code crash: \textit{Maximum call stack size exceeded}).}
  \label{tab:eval-detail-claude}
  \footnotesize
  \setlength{\tabcolsep}{3pt}
  \resizebox{0.93\linewidth}{!}{%
    \begin{tabular}{@{}l l c r r r r r@{}}
      \toprule
      Task & Model & Task Passed & Test-suite Pass Rate & Duration & Core LOC & Tokens (in/out) & Cost (\$)\\
      \midrule
      \multirow[c]{2}{*}{\texttt{pug}} & claude-opus-4.6 & No & 51.4\% (129/251) & 1.8h & 4246 & 44.57M/138.2k & 51.69\\
      & claude-opus-4.5 & No & 35.5\% (89/251) & 1.1h & 3422 & 30.96M/130.8k & 22.82\\
      \midrule
      \multirow[c]{2}{*}{\texttt{jq}} & claude-opus-4.6 & Yes & \textbf{100.0\% (218/218)} & 1.3h & 6055 & 27.74M/170.3k & 25.49\\
      & claude-opus-4.5 & Yes & \textbf{100.0\% (218/218)} & 1.3h & 7812 & 48.46M/245.9k & 34.47\\
      \midrule
      \multirow[c]{2}{*}{\texttt{csv}} & claude-opus-4.6 & Yes & \textbf{100.0\% (98/98)} & 0.70h & 474 & 16.27M/91.7k & 12.44\\
      & claude-opus-4.5 & Yes & \textbf{100.0\% (98/98)} & 0.49h & 483 & 15.54M/65.7k & 11.89\\
      \midrule
      \multirow[c]{2}{*}{\texttt{ini}} & claude-opus-4.6 & Yes & \textbf{100.0\% (98/98)} & 0.54h & 923 & 8.94M/80.7k & 9.05\\
      & claude-opus-4.5 & Yes & \textbf{100.0\% (98/98)} & 0.28h & 1070 & 8.28M/53.9k & 6.43\\
      \midrule
      \multirow[c]{2}{*}{\texttt{yaml}} & claude-opus-4.6 & No & 99.7\% (344/345) & 1.8h & 3721 & 42.91M/142.0k & 29.83\\
      & claude-opus-4.5 & No & 68.1\% (235/345) & 1.4h & 3596 & 46.98M/207.7k & 33.91\\
      \midrule
      \multirow[c]{2}{*}{\texttt{toml}} & claude-opus-4.6 & No & 98.0\% (718/733) & 13.0h & 6779 & N/A & N/A\\
      & claude-opus-4.5 & No & 82.3\% (603/733) & 0.65h & 2855 & 26.86M/106.9k & 18.27\\
      \midrule
      \multirow[c]{2}{*}{\texttt{xml}} & claude-opus-4.6 & Yes & \textbf{100.0\% (735/735)} & 1.6h & 2839 & 50.70M/172.3k & 35.88\\
      & claude-opus-4.5 & Yes & \textbf{100.0\% (735/735)} & 2.4h & 3241 & 58.68M/207.2k & 39.42\\
      \midrule
      \multirow[c]{2}{*}{\texttt{html5}} & claude-opus-4.6 & Yes & \textbf{100.0\% (8221/8221)} & 12.5h & 10585 & N/A & N/A\\
      & claude-opus-4.5 & No & 56.5\% (4648/8221) & 1.0h & 7583 & 17.39M/110.0k & 14.38\\
      \midrule
      \multirow[c]{2}{*}{\texttt{c99}} & claude-opus-4.6 & Yes & \textbf{100.0\% (117/117)} & 1.1h & 4979 & 23.48M/100.5k & 27.12\\
      & claude-opus-4.5 & No & 45.3\% (53/117) & 0.69h & 4937 & 30.44M/96.2k & 20.39\\
      \midrule
      \multirow[c]{2}{*}{\texttt{lua}} & claude-opus-4.6 & No & 97.1\% (133/137) & 1.5h & 6688 & 43.37M/155.4k & 398.24\\
      & claude-opus-4.5 & No & 96.4\% (132/137) & 1.2h & 6782 & 46.14M/220.0k & 34.15\\
      \midrule
      \multirow[c]{2}{*}{\texttt{ecma262}} & claude-opus-4.6 & No & 60.2\% (372/618) & 9.3h & 13901 & N/A & N/A\\
      & claude-opus-4.5 & No & 23.1\% (143/618) & 0.64h & 5172 & 34.13M/107.6k & 21.78\\
      \midrule
      \multirow[c]{2}{*}{\texttt{python}} & claude-opus-4.6 & No & 0.0\% (0/653) & 5.2h & 16991 & 127.99M/365.6k & 135.91\\
      & claude-opus-4.5 & No & 60.8\% (397/653) & 3.2h & 10932 & 84.85M/252.1k & 60.45\\
      \midrule
      \multirow[c]{2}{*}{\texttt{r6rs}} & claude-opus-4.6 & No & 91.9\% (1252/1362) & 7.9h & 20422 & 21.28M/199.5k & 190.32\\
      & claude-opus-4.5 & No & 54.0\% (735/1362) & 2.9h & 8585 & 51.90M/214.4k & 36.57\\
      \midrule
      \multirow[c]{2}{*}{\texttt{git\_object}} & claude-opus-4.6 & Yes & \textbf{100.0\% (1000/1000)} & 0.36h & 1291 & 6.79M/40.5k & 8.27\\
      & claude-opus-4.5 & Yes & \textbf{100.0\% (1000/1000)} & 0.68h & 1065 & 23.93M/66.9k & 15.81\\
      \midrule
      \multirow[c]{2}{*}{\texttt{protobuf}} & claude-opus-4.6 & Yes & \textbf{100.0\% (141/141)} & 0.20h & 858 & 6.45M/28.0k & 4.77\\
      & claude-opus-4.5 & Yes & \textbf{100.0\% (141/141)} & 0.17h & 1051 & 6.70M/32.4k & 5.21\\
      \midrule
      \multirow[c]{2}{*}{\texttt{zip}} & claude-opus-4.6 & Yes & \textbf{100.0\% (1089/1089)} & 8.1h & 10530 & 58.10M/227.1k & 1016.53\\
      & claude-opus-4.5 & No & 88.0\% (958/1089) & 2.1h & 2006 & 56.33M/174.9k & 37.19\\
      \midrule
      \multirow[c]{2}{*}{\texttt{capnp}} & claude-opus-4.6 & Yes & \textbf{100.0\% (111/111)} & 0.38h & 2605 & 10.63M/57.8k & 9.37\\
      & claude-opus-4.5 & Yes & \textbf{100.0\% (111/111)} & 0.25h & 2690 & 7.09M/47.6k & 7.55\\
      \midrule
      \multirow[c]{2}{*}{\texttt{wasm}} & claude-opus-4.6 & Yes & \textbf{100.0\% (800/800)} & 0.54h & 4152 & 16.50M/92.3k & 13.66\\
      & claude-opus-4.5 & Yes & \textbf{100.0\% (800/800)} & 1.7h & 6101 & 44.49M/164.6k & 30.91\\
      \midrule
      \multirow[c]{2}{*}{\texttt{uri}} & claude-opus-4.6 & Yes & \textbf{100.0\% (138/138)} & 0.23h & 1198 & 5.93M/39.6k & 4.72\\
      & claude-opus-4.5 & Yes & \textbf{100.0\% (138/138)} & 0.31h & 1320 & 8.01M/56.7k & 6.81\\
      \midrule
      \multirow[c]{2}{*}{\texttt{hpack}} & claude-opus-4.6 & Yes & \textbf{100.0\% (129/129)} & 0.68h & 5941 & 9.37M/63.3k & 9.37\\
      & claude-opus-4.5 & Yes & \textbf{100.0\% (129/129)} & 0.38h & 1561 & 10.84M/75.2k & 10.53\\
      \midrule
      \multirow[c]{2}{*}{\texttt{url}} & claude-opus-4.6 & Yes & \textbf{100.0\% (1220/1220)} & 1.2h & 4065 & 29.16M/126.3k & 26.97\\
      & claude-opus-4.5 & No & 87.0\% (1062/1220) & 1.1h & 2517 & 27.95M/94.5k & 18.36\\
      \midrule
      \multirow[c]{2}{*}{\texttt{cdcl}} & claude-opus-4.6 & Yes & \textbf{100.0\% (4312/4312)} & 6.7h & 1203 & 60.77M/221.0k & 46.19\\
      & claude-opus-4.5 & No & 99.6\% (4295/4312) & 3.0h & 1020 & 31.56M/82.9k & 20.63\\
      \bottomrule
    \end{tabular}
  }
\end{table}

\endgroup

\subsection{Main Results}
\label{sec:eval-results}

\paragraph{Overall performance.}
Table~\ref{tab:eval-summary} summarizes SWE-AGI performance by difficulty tier and reveals a sharp difficulty gradient.
On the easy tier, all evaluated frontier agents (gpt-5.3-codex, gpt-5.2-codex, claude-opus-4.6, claude-opus-4.5) solve 6/6 tasks with 100\% test-suite pass rate, indicating that for small parsers/decoders the end-to-end loop (spec reading, implementation under a fixed scaffold, and iteration under test feedback) can be executed reliably.
On the medium and hard tiers, outcomes diverge: gpt-5.3-codex solves 8/8 medium and 5/8 hard tasks (19/22 overall), gpt-5.2-codex solves 7/8 medium and 4/8 hard (17/22), claude-opus-4.6 solves 5/8 medium and 4/8 hard (15/22), while claude-opus-4.5 solves 3/8 medium and 1/8 hard (10/22).
This widening separation suggests that scaling to larger, more specification-intensive systems is the key differentiator among frontier agents in SWE-AGI.\@

\begin{sloppypar}
We also run a rapid easy-tier sweep of additional models.
Even within this easier regime, success rates are low.
kimi-k2.5, glm-4.7, and gemini-3-flash solve only 2/6 tasks.
deepseek-v3.2 solves 1/6, while claude-sonnet-4.5 and qwen3-max solve 0/6.
These results indicate that SWE-AGI is sensitive to robustness and generalization under specification pressure: models that appear close on code-centric open benchmarks can separate substantially once placed in an end-to-end setting with hidden private tests.
\end{sloppypar}

Failure to solve a task does not always indicate broad functional incorrectness.
Across tiers, many ``failed'' submissions still pass a large fraction of the evaluation test suite, suggesting that remaining defects are often localized to rare normative requirements, subtle state-machine corner cases, or performance bottlenecks that only surface in the hidden private tests.
This is most pronounced on the hard tier: despite solving fewer hard tasks than gpt-5.3-codex (4/8 vs.\ 5/8), gpt-5.2-codex achieves a higher unweighted mean hard-tier test-suite pass rate (91.2\%), reflecting near-complete coverage on several failures.
At the task level, we observe multiple near-misses, e.g., \texttt{cdcl} reaches 99.8\% test-suite pass rate for gpt-5.2-codex and \texttt{lua} reaches 96.4\% for claude-opus-4.5 (Tables~\ref{tab:eval-detail-gpt} and~\ref{tab:eval-detail-claude}).
Practically, this means the pass/fail boundary is often dominated by eliminating the last few spec-sensitive edge cases rather than constructing missing core subsystems.

\paragraph{Agent Efficiency.}
Average wall-clock time and code size primarily reflect long-horizon engineering difficulty and agent efficiency bottlenecks, rather than pure model capability; both are also strongly influenced by the chosen front-end configuration and tool policies, and should therefore be interpreted with caution.
Within this framing, Table~\ref{tab:eval-summary} highlights two consistent gaps.
First, gpt-5.3-codex is substantially more time-efficient than gpt-5.2-codex while also improving task completion: its average runtime is about 3--5$\times$ lower across tiers (0.28h vs.\ 0.81h on easy, 1.2h vs.\ 5.1h on medium, 1.7h vs.\ 7.8h on hard), and its average implementations are smaller on medium and hard tasks (2575 vs.\ 4702 core LOC on medium; 6255 vs.\ 9034 on hard).
Second, claude-opus-4.6 improves substantially over claude-opus-4.5 on medium and hard tiers (15/22 vs.\ 10/22 overall), but this gain comes with higher wall-clock time on those tiers (3.5h vs.\ 1.3h on medium; 5.7h vs.\ 1.7h on hard), consistent with additional exploration and debugging under specification pressure.

At the same time, the runs reveal a noteworthy capability of gpt-5.2-codex: sustained long-horizon execution even when convergence fails.
For example, on \texttt{ecma262} the agent runs for 42 hours without early termination while still failing the private test suite, producing an unusually large implementation (over 30k core LOC).
Accordingly, we treat core LOC as a coarse indicator of implementation scale rather than an optimization target: higher LOC may indicate broader feature coverage, but may also reflect verbose implementations and refactoring churn under heavy specification pressure.

\FloatBarrier

\begin{table}[t]
  \centering
  \caption{SWE behavior categories used for log-based analysis. Categories are heuristic labels applied to logged tool actions to summarize effort allocation.}
  \label{tab:behavior-definitions}
  \footnotesize
  \setlength{\tabcolsep}{3pt}
  \renewcommand{\arraystretch}{1.08}
  \begin{tabular}{@{}l l p{10.6cm}@{}}
    \toprule
    Abbrev. & Category & Definition (typical signals)\\
    \midrule
    \textbf{Action} & Action count & Counted logged tool actions for a task run (used as a coarse proxy for interaction volume).\\
    \textbf{Spec} & Spec understanding & Reading/searching requirements and expected behavior (e.g., \texttt{TASK.md}, \texttt{specs/}, public tests).\\
    \textbf{Plan} & Planning & Creating or updating task-level plans/todo items.\\
    \textbf{Read} & Code understanding & Reading/searching implementation code or inspecting artifacts to understand implementation; includes repository exploration and file navigation (e.g., \texttt{ls}, \texttt{find}, help).\\
    \textbf{Write} & Code writing & Creating/modifying project files (implementation/config), including edits and file operations.\\
    \textbf{Debug} & Debugging & Running builds/tests/submissions and investigating failures; includes non-zero-exit actions.\\
    \textbf{Hyg} & Hygiene & Formatting or mechanical refactors explicitly recorded (e.g., \texttt{moon fmt}).\\
    \textbf{Ext} & External search & Fetching/searching resources outside the repository.\\
    \textbf{Other} & Other & Remaining actions that do not clearly fit the above categories.\\
    \bottomrule
  \end{tabular}
\end{table}

\subsection{End-to-End SWE Behavior Analysis}
\label{sec:eval-behavior}
Beyond pass/fail outcomes, we analyze how agents allocate effort over long trajectories by labeling logged tool actions into coarse SWE-relevant behavior categories.
The taxonomy (Table~\ref{tab:behavior-definitions}) is heuristic: it maps observable actions (shell commands, file reads/writes, test runs, submissions, etc.) to a small set of intent-level buckets that approximate the engineering loop (spec understanding, code understanding/writing, debugging, hygiene, and external search).
These statistics do not capture unlogged internal reasoning, and absolute counts depend on each agent front-end's logging granularity; we therefore interpret them as qualitative indicators of \emph{effort allocation} rather than a normalized efficiency metric.

\begin{table}[!htbp]
  \centering
  \caption{Behavior summary by difficulty tier (percent of logged actions). Action reports average counted actions per task. Top-3 behavior shares per row are bold.}
  \label{tab:behavior-tier}
  \footnotesize
  \setlength{\tabcolsep}{3pt}
  \renewcommand{\arraystretch}{0.95}{
    \begin{tabular}{@{}clrrrrrrrrr@{}}
      \toprule
      Difficulty & Model & Action & Spec & Plan & Read & Write & Debug & Hyg & Ext & Other\\
      \midrule
      \multirow[c]{10}{*}{Easy} & gpt-5.3-codex & 109 & 9.3\% & 0.2\% & \textbf{46.0\%} & \textbf{24.8\%} & \textbf{17.7\%} & 0.5\% & 0.0\% & 1.5\% \\
      & gpt-5.2-codex & 195 & 6.8\% & 0.4\% & \textbf{50.1\%} & \textbf{25.3\%} & \textbf{12.1\%} & 3.4\% & 0.0\% & 1.9\% \\
      & claude-opus-4.6 & 150 & 11.9\% & 6.6\% & \textbf{35.1\%} & \textbf{24.1\%} & \textbf{17.4\%} & 0.3\% & 1.2\% & 3.3\% \\
      & claude-opus-4.5 & 145 & 14.9\% & 6.6\% & \textbf{25.3\%} & \textbf{27.0\%} & \textbf{24.0\%} & 0.1\% & 0.3\% & 1.8\% \\
      & claude-sonnet-4.5 & 165 & 5.3\% & 5.5\% & \textbf{25.8\%} & \textbf{35.9\%} & \textbf{24.1\%} & 0.4\% & 0.0\% & 3.1\% \\
      & kimi-k2.5 & 138 & 6.7\% & 3.3\% & \textbf{20.3\%} & \textbf{26.8\%} & \textbf{33.7\%} & 0.5\% & 0.0\% & 8.8\% \\
      & glm-4.7 & 232 & 4.7\% & 3.7\% & \textbf{30.8\%} & \textbf{26.3\%} & \textbf{27.9\%} & 0.1\% & 1.1\% & 5.3\% \\
      & gemini-3-flash & 79 & 5.9\% & 0.0\% & \textbf{12.9\%} & \textbf{46.2\%} & \textbf{33.7\%} & 0.0\% & 0.2\% & 1.1\% \\
      & deepseek-v3.2 & 608 & 4.2\% & 3.3\% & \textbf{36.2\%} & \textbf{38.4\%} & \textbf{17.1\%} & 0.1\% & 0.2\% & 0.4\% \\
      & qwen3-max & 198 & 7.1\% & 4.4\% & \textbf{24.5\%} & \textbf{42.4\%} & \textbf{21.2\%} & 0.0\% & 0.4\% & 0.1\% \\
      \midrule
      \multirow[c]{4}{*}{Medium} & gpt-5.3-codex & 311 & 7.2\% & 0.0\% & \textbf{39.6\%} & \textbf{19.3\%} & \textbf{19.1\%} & 0.8\% & 6.0\% & 7.9\% \\
      & gpt-5.2-codex & 1070 & 6.1\% & 0.0\% & \textbf{50.1\%} & \textbf{21.1\%} & \textbf{11.6\%} & 1.9\% & 2.0\% & 7.2\% \\
      & claude-opus-4.6 & 938 & 8.2\% & 5.7\% & \textbf{43.1\%} & \textbf{13.8\%} & \textbf{14.2\%} & 0.1\% & 10.2\% & 4.6\% \\
      & claude-opus-4.5 & 381 & 6.2\% & 3.6\% & \textbf{35.9\%} & \textbf{25.8\%} & \textbf{24.9\%} & 0.2\% & 0.2\% & 3.1\% \\
      \midrule
      \multirow[c]{4}{*}{Hard} & gpt-5.3-codex & 301 & 6.0\% & 0.0\% & \textbf{41.4\%} & \textbf{20.9\%} & \textbf{19.8\%} & 0.5\% & 2.0\% & 9.3\% \\
      & gpt-5.2-codex & 1676 & 2.3\% & 0.0\% & \textbf{64.6\%} & \textbf{20.5\%} & \textbf{9.2\%} & 1.0\% & 0.2\% & 2.1\% \\
      & claude-opus-4.6 & 1498 & 6.6\% & 6.7\% & \textbf{50.2\%} & \textbf{13.3\%} & \textbf{16.2\%} & 0.1\% & 3.4\% & 3.5\% \\
      & claude-opus-4.5 & 434 & 5.2\% & 4.4\% & \textbf{43.5\%} & \textbf{24.5\%} & \textbf{20.3\%} & 0.1\% & 0.2\% & 1.8\% \\
      \bottomrule
    \end{tabular}%
  }
\end{table}

Table~\ref{tab:behavior-tier} summarizes the distribution of agent behaviors across difficulty tiers.
As difficulty increases, code understanding (Read) becomes the dominant activity and interaction volume grows sharply for several agents.
On hard tasks, Read accounts for 41.4\% of logged actions for gpt-5.3-codex and 64.6\% for gpt-5.2-codex, with claude-opus-4.6 at 50.2\% and claude-opus-4.5 at 43.5\%.
This shift coincides with a large increase in total actions: on hard tasks, gpt-5.2-codex averages 1676 logged actions per task, compared to 301 for gpt-5.3-codex and 1498 for claude-opus-4.6.
Overall, once implementations reach multi-module, spec-heavy regimes, agents devote a substantial fraction of their effort to reading, inspecting, and validating existing code rather than generating new functionality.

These patterns suggest that long-horizon progress is constrained less by raw code generation capacity than by the ability to maintain and reason over an evolving codebase.
In this setting the bottleneck shifts toward preserving architectural consistency, understanding prior design decisions, and verifying interactions across modules.
This aligns with findings in~\cite{thomas2026darkflow} that identify code reading---rather than code writing---as a central bottleneck in AI-assisted software development, and supports the view that comprehension and maintenance costs dominate long-horizon engineering.

\paragraph{Strategy Differences Across Frontier Agents.}
Frontier agents exhibit systematic differences in workflow that track within-family improvements.
Relative to gpt-5.2-codex, gpt-5.3-codex is markedly more iteration-oriented on medium and hard tasks: it spends a smaller share on Read (41.4\% vs.\ 64.6\% on hard) while allocating more to Debug (19.8\% vs.\ 9.2\%), and it completes runs with far fewer logged actions (301 vs.\ 1676 on hard).
This profile is consistent with faster convergence: fewer prolonged ``maintenance'' phases dominated by reading and more decisive test--fix--retest loops, yielding substantially lower wall-clock time while improving task completion (Table~\ref{tab:eval-summary}).

Within the Claude family, claude-opus-4.6 improves substantially over claude-opus-4.5 on medium and hard tiers, and its behavior suggests a more deliberate workflow.
Compared to claude-opus-4.5, it allocates more effort to specification engagement and planning (e.g., on hard tasks: 6.6\% Spec and 6.7\% Plan vs.\ 5.2\% Spec and 4.4\% Plan) and less to raw code writing (13.3\% vs.\ 24.5\%), while maintaining a comparable debugging share (16.2\% vs.\ 20.3\%).
This shift toward reading and planning appears beneficial on spec-heavy systems where naive patching can destabilize global invariants.
In contrast, claude-opus-4.5 exhibits a more pronounced ``read specification--patch--rerun'' pattern, with higher Write and Debug shares across tiers.
While such a strategy can be effective on smaller tasks where localized fixes converge quickly, on complex state-machine--driven systems (e.g., the HTML5 parser) frequent local patches may accumulate inconsistencies and degrade architectural coherence, leading to instability rather than convergence.

\section{Related Work}

\paragraph{Evaluation of LLMs.}
Broad evaluation frameworks such as HELM \citep{liang2022helm} and BIG-bench \citep{srivastava2022bigbench} emphasize multi-scenario, multi-metric measurement, highlighting trade-offs beyond accuracy, such as robustness and efficiency. As LLMs increasingly transition into autonomous agents \citep{yao2022react,schick2023toolformer,shinn2023reflexion}, evaluation has shifted from static prompting to interactive environments that stress tool use, multi-step planning, and long-horizon consistency. While domain-agnostic benchmarks like AgentBench \citep{liu2023agentbench} and Terminal-Bench \citep{tbench2025terminalbench} provide foundational infrastructure, SWE-AGI focuses on the unique constraints of software engineering. It departs from the repository-centric paradigm of SWE-bench \citep{jimenez2023swebench} in two key ways: (i) tasks are defined by rigorous, ground-truth specifications rather than existing codebase conventions, and (ii) it employs a submission-based sandbox with private, non-public test suites, ensuring auditable measurement even for models with unrestricted web search and retrieval capabilities.

\paragraph{Software Engineering Benchmarks.}
The evaluation of code intelligence has evolved from snippet-level synthesis to full-lifecycle engineering. Early benchmarks like HumanEval \citep{chen2021evaluating} and MBPP \citep{austin2021program} focus on isolated function-level tasks, while efforts like EvalPlus \citep{liu2023evalplus} address test-case insufficiency. To counter data contamination, LiveCodeBench \citep{jain2024livecodebench} introduced continuous curation. However, real-world engineering requires reasoning across multiple files, as explored in RepoBench \citep{liu2023repobench} and SWE-bench \citep{jimenez2023swebench}. Recently, the design space has expanded toward specialized dimensions: PRDBench \citep{fu2025automaticallybenchmarkingllmcode} targets PRD-to-code workflows; OSS-Bench \citep{jiang2025ossbenchbenchmarkgeneratorcoding} focuses on memory-safety and optimization; and SWE-EVO \citep{thai2026sweevobenchmarkingcodingagents} shifts from initial construction to continuous software evolution. SWE-AGI complements this landscape by targeting the end-to-end systems regime: agents must build a complete, robust system from high-level specs under a fixed API.\@ By decoupling the evaluation from visible unit tests and existing repository noise, SWE-AGI provides a cleaner signal for an agent's ability to handle the ``requirements-to-implementation'' gap---a critical frontier for production-scale AI engineering.

\paragraph{Programming Languages and LLMs.}
Programming languages and ecosystems shape what models can learn and how reliably they generalize. MultiPL-E \citep{cassano2022multiplE} shows that model performance and failure modes vary across languages, reflecting differences in syntax, standard libraries, tooling, and conventions. Beyond syntax, effective AI coding increasingly depends on a ``full-stack'' tool-and-feedback loop: editor/refactoring support, build systems, test runners, linters, static analyzers, profilers, and submission/evaluation harnesses that provide fast and accurate signals. In many real deployments, the bottleneck is not code generation but review, debugging, integration, and specification clarification---suggesting an advantage for languages and platforms that shift feedback from humans to machines via strong static guarantees, deterministic builds, and rich automated checks.

This favors statically typed languages and ecosystems that integrate a one-stop toolchain and enforce disciplined interfaces, enabling agents to iterate with high-quality feedback and fewer ambiguous failure modes. As the fraction of AI-generated code grows, language and platform design may increasingly optimize for machine-assisted development: explicit specifications, stable API scaffolds, auditable build/test pipelines, and standard diagnostics that can be consumed by agents. SWE-AGI uses MoonBit \citep{moonbit}, a recently developed programming language with an integrated toolchain: the \texttt{declare} keyword supports declaration-first scaffolding under a fixed API, and the unified workflow (\texttt{moon}) supports fast compilation, reproducible builds, and submission-style evaluation at production scale.

\section{Conclusion}

SWE-AGI evaluates LLM-based software engineering agents on tasks defined by explicit specifications and measured by deterministic, human-validated tests.
The benchmark targets production-quality, from-scratch MoonBit implementations in the $10^3$--$10^4$~LOC regime and is evaluated through an iterative submission protocol: agents build and test locally, submit via \texttt{swe-agi-submit}, and receive pass/fail feedback from hidden private tests.
Across 22 tasks spanning seven specification families, we observe a steep difficulty gradient: frontier agents reliably solve all easy tasks, but performance drops sharply on medium and hard tiers.
Overall, {\small gpt-5.3-codex} solves 19/22 tasks (86.4\%), {\small gpt-5.2-codex} solves 17/22 (77.3\%), {\small claude-opus-4.6} solves 15/22 (68.2\%), and {\small claude-opus-4.5} solves 10/22 (45.5\%).
Many failures are near-misses with high test-suite pass rates, suggesting that the pass/fail boundary is often dominated by a small number of specification-sensitive edge cases and performance corner cases rather than missing major subsystems.

Complementing these outcome metrics, our log-based behavior analysis indicates that long-horizon progress is increasingly dominated by code understanding and maintenance rather than raw code writing.
As difficulty increases, agents spend a growing share of actions reading and inspecting evolving implementations, and systematic differences in Read/Write/Debug allocation track within-family performance improvements.
These findings reinforce that the central bottleneck in end-to-end agentic software engineering is sustaining coherent, correct systems over long trajectories under build/test feedback.

In future work, we will extend SWE-AGI to encompass heterogeneous distributed systems and complex legacy code integration tasks that demand deep architectural reasoning.
We also plan to study library-centric workflows: how agents decompose specifications into reusable components, divide subtasks across libraries, and compose existing libraries into even larger software systems.
Finally, incorporating multi-modal inputs (e.g., architectural diagrams and visual execution traces) and exploring agent-centric toolchain optimizations alongside non-functional imperatives like security and maintainability will be essential for achieving deterministic, production-grade reliability.

\FloatBarrier
\begingroup
\footnotesize
\bibliography{main}
\endgroup

\newpage

\appendix

\section{SWE-AGI Task Suite}
\label{app:tasks}

\begingroup
\scriptsize
\setlength{\tabcolsep}{3pt}
\renewcommand{\arraystretch}{1.08}
\begin{longtable}{@{}>{\raggedright\arraybackslash}p{5.4cm} >{\centering\arraybackslash}p{1.2cm} >{\raggedleft\arraybackslash}p{1.6cm} >{\raggedright\arraybackslash}p{6.3cm}@{}}
  \caption{SWE-AGI task suite (22 tasks, 7 categories). Core LOC (excluding tests and tooling) is reported as a coarse magnitude estimate (derived from benchmarked agent implementations and excluding public and private tests). Rows are sorted by core LOC within each category.}
  \label{tab:tasks}\\
  \toprule
  Task (\texttt{id}: title) & Difficulty & Core LOC & Key complexity drivers \\
  \midrule
  \endfirsthead

  \multicolumn{4}{@{}l}{\tablename~\thetable\ (continued).}\\
  \toprule
  Task (\texttt{id}: title) & Difficulty & Core LOC & Key complexity drivers \\
  \midrule
  \endhead

  \midrule
  \multicolumn{4}{r@{}}{\scriptsize Continued on next page.}\\
  \endfoot

  \bottomrule
  \endlastfoot

  \multicolumn{4}{@{}l}{\textbf{Totals:} 22 tasks (Easy=6, Medium=8, Hard=8).} \\
  \addlinespace[2pt]
  \multicolumn{4}{@{}l}{\textbf{Template and Domain-Specific Languages}} \\
  \texttt{pug}: Pug Template Language & Medium & $\sim 5\times 10^3$ & Indentation semantics, mixins/blocks, scope/inclusion, error localization \\
  \texttt{jq}: JQ Query Language Interpreter & Hard & $\sim 7\times 10^3$ & Lexer/parser, stream semantics (0..N outputs), built-ins, error modes \\
  \addlinespace[2pt]
  \multicolumn{4}{@{}l}{\textbf{Data Serialization and Configuration Formats}} \\
  \texttt{csv}: CSV Parser (RFC 4180) & Easy & $\sim 10^3$ & Quoting/escaping, multiline fields, line ending edge cases, invalid patterns \\
  \texttt{ini}: INI Parser & Easy & $\sim 10^3$ & Section/key parsing, escaping rules, normalization, error handling \\
  \texttt{yaml}: YAML 1.2 Parser & Medium & $\sim 3\times 10^3$ & Indentation/block structure, anchors/tags, scalars, error recovery \\
  \texttt{toml}: TOML 1.0 Parser & Medium & $\sim 3\times 10^3$ & Dotted keys, array-of-tables, datetime/float rules, UTF-8 + diagnostics \\
  \addlinespace[2pt]
  \multicolumn{4}{@{}l}{\textbf{Markup and Document Formats}} \\
  \texttt{xml}: XML 1.0 + Namespaces & Medium & $\sim 3\times 10^3$ & Well-formedness, namespaces, entities/DTD subset, error handling, streaming/DOM tradeoffs \\
  \texttt{html5}: HTML5 Parser & Hard & $\sim 10^4$ & Tokenization + tree builder state machines, error recovery, entities, broad conformance \\
  \addlinespace[2pt]
  \multicolumn{4}{@{}l}{\textbf{Programming Language Front-Ends}} \\
  \texttt{c99}: C99 Parser & Hard & $\sim 5\times 10^3$ & Declarators/type system, precedence/ambiguity, AST + symbols, error recovery \\
  \texttt{lua}: Lua 5.4 Interpreter & Hard & $\sim 5\times 10^3$ & VM/bytecode, tables + metatables, closures, coroutines, GC scope \\
  \texttt{ecma262}: ECMAScript Interpreter (ECMA-262 subset) & Hard & $\sim 7\times 10^3$ & Parsing + semantics, runtime objects, corner cases exercised by suite \\
  \texttt{python}: Python Interpreter (subset) & Hard & $\sim 7\times 10^3$ & Indentation lexing, object model, exceptions, scoping/closures, built-ins \\
  \texttt{r6rs}: R6RS Scheme Interpreter (subset) & Hard & $\sim 7\times 10^3$ & Reader, macro system, evaluator/runtime, exact printing semantics \\
  \addlinespace[2pt]
  \multicolumn{4}{@{}l}{\textbf{Binary Formats and Streaming Decoders}} \\
  \texttt{git\_object}: Git Object Parser (loose objects) & Easy & $\sim 10^3$ & zlib integration, header parsing, hashing, boundary/error handling \\
  \texttt{protobuf}: Protocol Buffers (streaming codec) & Easy & $\sim 10^3$ & Varint/zigzag, length-delimited fields, chunked reads, malformed input handling \\
  \texttt{zip}: ZIP File Parser & Medium & $\sim 3\times 10^3$ & Central directory, Zip64, streaming reads, CRC/validation, encoding details \\
  \texttt{capnp}: Cap'n Proto Binary Format & Medium & $\sim 3\times 10^3$ & Packed encoding, pointers/segments, far pointers, boundary safety \\
  \texttt{wasm}: WASM Decoder + Validator & Medium & $\sim 5\times 10^3$ & LEB128, section/index consistency, validation rules, precise error behavior \\
  \addlinespace[2pt]
  \multicolumn{4}{@{}l}{\textbf{Networking and Protocol State Machines}} \\
  \texttt{uri}: URI Parser (RFC 3986) & Easy & $\sim 10^3$ & Normalization and resolution rules, encoding constraints, error behavior \\
  \texttt{hpack}: HPACK Decoder/Encoder (RFC 7541) & Easy & $\sim 10^3$ & Huffman coding, dynamic table management, header field semantics \\
  \texttt{url}: URL Parser (WHATWG) & Medium & $\sim 3\times 10^3$ & Canonicalization, relative resolution, percent-encoding, IDNA/Punycode scope \\
  \addlinespace[2pt]
  \multicolumn{4}{@{}l}{\textbf{Automated Reasoning and SAT Solving}} \\
  \texttt{cdcl}: CDCL SAT Solver & Hard & $\sim 2\times 10^3$ & Unit propagation, clause learning, backtracking/heuristics, data-structure efficiency \\
\end{longtable}
\endgroup

\section{Detailed Results on SWE Behaviors}
\label{app:behavior-tables}
Table~\ref{tab:behavior-frontier-gpt} collects the per-task behavior stats tables referenced in Section~\ref{sec:eval-behavior}: the first part reports gpt-5.3-codex and gpt-5.2-codex, and the continuation reports claude-opus-4.6 and claude-opus-4.5.
Percentages denote the share of logged tool actions assigned to each behavior category (Spec, Plan, Read, Write, Debug, Hyg, Ext, Other); for readability, the top-3 behavior shares per row are bold.
In Table~\ref{tab:behavior-frontier-gpt}, the \emph{Action} column reports the counted logged actions for that task run and should be interpreted as a coarse proxy for interaction volume rather than a normalized efficiency measure, since logging granularity varies across agent front-ends and runs.

\begin{table}[!htbp]
\centering
\caption{Per-task behavior stats for gpt-5.3-codex and gpt-5.2-codex (percent of logged actions). Top-3 behavior shares per row are bold.}
\label{tab:behavior-frontier-gpt}
\begingroup
\scriptsize
\setlength{\tabcolsep}{2pt}
\renewcommand{\arraystretch}{0.92}
\resizebox{0.90\linewidth}{!}{%
\begin{tabular}{@{}l l r r r r r r r r r@{}}
\toprule
Task & Model & Action & Spec & Plan & Read & Write & Debug & Hyg & Ext & Other\\
\midrule
\multirow[c]{2}{*}{\texttt{pug}} & gpt-5.3-codex & 708 & 5.5\% & 0.0\% & \textbf{35.6\%} & \textbf{18.4\%} & \textbf{15.0\%} & 1.0\% & 13.6\% & 11.0\% \\
 & gpt-5.2-codex & 5093 & 5.4\% & 0.0\% & \textbf{54.0\%} & \textbf{19.2\%} & \textbf{10.3\%} & 1.3\% & 2.7\% & 7.0\% \\
\midrule
\multirow[c]{2}{*}{\texttt{jq}} & gpt-5.3-codex & 435 & 2.8\% & 0.0\% & \textbf{32.0\%} & 20.7\% & \textbf{22.5\%} & 0.2\% & 0.0\% & \textbf{21.8\%} \\
 & gpt-5.2-codex & 522 & 3.3\% & 0.0\% & \textbf{45.2\%} & \textbf{31.6\%} & \textbf{16.7\%} & 3.3\% & 0.0\% & 0.0\% \\
\midrule
\multirow[c]{2}{*}{\texttt{csv}} & gpt-5.3-codex & 84 & 8.3\% & 0.0\% & \textbf{48.8\%} & \textbf{22.6\%} & \textbf{17.9\%} & 0.0\% & 0.0\% & 2.4\% \\
 & gpt-5.2-codex & 143 & 7.7\% & 0.7\% & \textbf{56.6\%} & \textbf{19.6\%} & \textbf{9.1\%} & 5.6\% & 0.0\% & 0.7\% \\
\midrule
\multirow[c]{2}{*}{\texttt{ini}} & gpt-5.3-codex & 99 & 5.1\% & 0.0\% & \textbf{40.4\%} & \textbf{29.3\%} & \textbf{25.3\%} & 0.0\% & 0.0\% & 0.0\% \\
 & gpt-5.2-codex & 176 & 4.0\% & 0.0\% & \textbf{42.6\%} & \textbf{31.8\%} & \textbf{15.9\%} & 5.1\% & 0.0\% & 0.6\% \\
\midrule
\multirow[c]{2}{*}{\texttt{yaml}} & gpt-5.3-codex & 288 & 6.6\% & 0.0\% & \textbf{50.0\%} & \textbf{19.8\%} & \textbf{17.7\%} & 0.3\% & 0.7\% & 4.9\% \\
 & gpt-5.2-codex & 721 & 3.5\% & 0.1\% & \textbf{40.9\%} & \textbf{34.0\%} & \textbf{16.0\%} & 3.6\% & 0.8\% & 1.1\% \\
\midrule
\multirow[c]{2}{*}{\texttt{toml}} & gpt-5.3-codex & 237 & 12.2\% & 0.0\% & \textbf{33.3\%} & \textbf{14.3\%} & \textbf{30.4\%} & 3.4\% & 5.5\% & 0.8\% \\
 & gpt-5.2-codex & 474 & 8.9\% & 0.0\% & \textbf{39.0\%} & \textbf{28.1\%} & \textbf{19.2\%} & 4.0\% & 0.0\% & 0.8\% \\
\midrule
\multirow[c]{2}{*}{\texttt{xml}} & gpt-5.3-codex & 339 & 5.9\% & 0.0\% & \textbf{30.1\%} & \textbf{32.2\%} & \textbf{22.4\%} & 0.0\% & 8.0\% & 1.5\% \\
 & gpt-5.2-codex & 483 & 4.1\% & 0.2\% & \textbf{42.2\%} & \textbf{32.5\%} & \textbf{16.4\%} & 4.6\% & 0.0\% & 0.0\% \\
\midrule
\multirow[c]{2}{*}{\texttt{html5}} & gpt-5.3-codex & 329 & 7.9\% & 0.0\% & \textbf{30.1\%} & \textbf{20.1\%} & \textbf{29.5\%} & 0.6\% & 7.3\% & 4.6\% \\
 & gpt-5.2-codex & 386 & 10.1\% & 0.3\% & \textbf{33.9\%} & \textbf{28.2\%} & \textbf{21.0\%} & 0.5\% & 0.0\% & 6.0\% \\
\midrule
\multirow[c]{2}{*}{\texttt{c99}} & gpt-5.3-codex & 218 & 6.9\% & 0.0\% & \textbf{39.9\%} & \textbf{25.2\%} & \textbf{19.7\%} & 0.5\% & 0.5\% & 7.3\% \\
 & gpt-5.2-codex & 580 & 10.9\% & 0.2\% & \textbf{48.6\%} & \textbf{25.7\%} & \textbf{11.9\%} & 2.4\% & 0.0\% & 0.3\% \\
\midrule
\multirow[c]{2}{*}{\texttt{lua}} & gpt-5.3-codex & 215 & 4.2\% & 0.0\% & \textbf{42.3\%} & \textbf{35.8\%} & \textbf{16.7\%} & 0.0\% & 0.0\% & 0.9\% \\
 & gpt-5.2-codex & 568 & 1.8\% & 0.2\% & \textbf{52.6\%} & \textbf{34.7\%} & \textbf{9.3\%} & 1.1\% & 0.0\% & 0.4\% \\
\midrule
\multirow[c]{2}{*}{\texttt{ecma262}} & gpt-5.3-codex & 201 & 10.0\% & 0.0\% & \textbf{48.3\%} & \textbf{21.9\%} & \textbf{14.9\%} & 0.5\% & 2.5\% & 2.0\% \\
 & gpt-5.2-codex & 9794 & 0.8\% & 0.0\% & \textbf{71.2\%} & \textbf{17.3\%} & \textbf{7.8\%} & 0.4\% & 0.3\% & 2.1\% \\
\midrule
\multirow[c]{2}{*}{\texttt{python}} & gpt-5.3-codex & 419 & 5.3\% & 0.0\% & \textbf{56.1\%} & \textbf{20.3\%} & \textbf{12.6\%} & 0.2\% & 2.9\% & 2.6\% \\
 & gpt-5.2-codex & 511 & 3.1\% & 0.0\% & \textbf{55.4\%} & \textbf{24.9\%} & \textbf{10.2\%} & 1.6\% & 0.0\% & 4.9\% \\
\midrule
\multirow[c]{2}{*}{\texttt{r6rs}} & gpt-5.3-codex & 377 & 8.2\% & 0.0\% & \textbf{46.2\%} & 12.5\% & \textbf{15.9\%} & 1.6\% & 1.6\% & \textbf{14.1\%} \\
 & gpt-5.2-codex & 654 & \textbf{10.4\%} & 0.2\% & \textbf{49.1\%} & \textbf{28.9\%} & 10.2\% & 0.2\% & 0.0\% & 1.1\% \\
\midrule
\multirow[c]{2}{*}{\texttt{git\_object}} & gpt-5.3-codex & 125 & 4.0\% & 0.0\% & \textbf{59.2\%} & \textbf{24.8\%} & \textbf{9.6\%} & 0.8\% & 0.0\% & 1.6\% \\
 & gpt-5.2-codex & 324 & 4.6\% & 0.3\% & \textbf{54.3\%} & \textbf{23.5\%} & \textbf{12.7\%} & 3.4\% & 0.0\% & 1.2\% \\
\midrule
\multirow[c]{2}{*}{\texttt{protobuf}} & gpt-5.3-codex & 109 & \textbf{13.8\%} & 0.0\% & \textbf{46.8\%} & \textbf{24.8\%} & 12.8\% & 1.8\% & 0.0\% & 0.0\% \\
 & gpt-5.2-codex & 136 & \textbf{12.5\%} & 0.7\% & \textbf{54.4\%} & \textbf{20.6\%} & 8.1\% & 2.2\% & 0.0\% & 1.5\% \\
\midrule
\multirow[c]{2}{*}{\texttt{zip}} & gpt-5.3-codex & 128 & 5.5\% & 0.0\% & \textbf{52.3\%} & \textbf{22.7\%} & \textbf{12.5\%} & 0.0\% & 0.0\% & 7.0\% \\
 & gpt-5.2-codex & 693 & 2.5\% & 0.0\% & \textbf{42.6\%} & \textbf{12.0\%} & 7.8\% & 1.2\% & 1.9\% & \textbf{32.2\%} \\
\midrule
\multirow[c]{2}{*}{\texttt{capnp}} & gpt-5.3-codex & 194 & 13.4\% & 0.0\% & \textbf{42.8\%} & \textbf{19.1\%} & \textbf{21.6\%} & 1.0\% & 0.0\% & 2.1\% \\
 & gpt-5.2-codex & 265 & \textbf{12.5\%} & 0.0\% & \textbf{55.8\%} & \textbf{18.1\%} & 10.9\% & 1.9\% & 0.0\% & 0.8\% \\
\midrule
\multirow[c]{2}{*}{\texttt{wasm}} & gpt-5.3-codex & 263 & 10.3\% & 0.0\% & \textbf{42.6\%} & \textbf{14.8\%} & \textbf{16.3\%} & 0.0\% & 4.2\% & 11.8\% \\
 & gpt-5.2-codex & 491 & 10.4\% & 0.2\% & \textbf{49.1\%} & \textbf{21.0\%} & \textbf{13.4\%} & 2.4\% & 3.3\% & 0.2\% \\
\midrule
\multirow[c]{2}{*}{\texttt{uri}} & gpt-5.3-codex & 95 & 12.6\% & 0.0\% & \textbf{33.7\%} & \textbf{25.3\%} & \textbf{28.4\%} & 0.0\% & 0.0\% & 0.0\% \\
 & gpt-5.2-codex & 187 & 4.3\% & 0.5\% & \textbf{49.2\%} & \textbf{28.9\%} & \textbf{14.4\%} & 2.1\% & 0.0\% & 0.5\% \\
\midrule
\multirow[c]{2}{*}{\texttt{hpack}} & gpt-5.3-codex & 142 & 12.0\% & 0.7\% & \textbf{44.4\%} & \textbf{22.5\%} & \textbf{16.2\%} & 0.0\% & 0.0\% & 4.2\% \\
 & gpt-5.2-codex & 204 & \textbf{10.8\%} & 0.5\% & \textbf{43.1\%} & \textbf{26.5\%} & 10.3\% & 2.5\% & 0.0\% & 6.4\% \\
\midrule
\multirow[c]{2}{*}{\texttt{url}} & gpt-5.3-codex & 125 & 7.2\% & 0.0\% & \textbf{53.6\%} & \textbf{16.0\%} & \textbf{17.6\%} & 0.8\% & 0.0\% & 4.8\% \\
 & gpt-5.2-codex & 338 & \textbf{17.8\%} & 0.0\% & \textbf{48.5\%} & \textbf{16.6\%} & 10.4\% & 0.6\% & 0.0\% & 6.2\% \\
\midrule
\multirow[c]{2}{*}{\texttt{cdcl}} & gpt-5.3-codex & 216 & 4.2\% & 0.0\% & \textbf{35.2\%} & \textbf{18.5\%} & \textbf{28.2\%} & 0.5\% & 0.0\% & 13.4\% \\
 & gpt-5.2-codex & 393 & 2.3\% & 0.3\% & \textbf{35.9\%} & \textbf{31.3\%} & \textbf{14.2\%} & 13.5\% & 0.0\% & 2.5\% \\
\bottomrule
\end{tabular}%
}
\endgroup
\end{table}

\begin{table}[!htbp]
\ContinuedFloat
\centering
\caption{Per-task behavior stats (continued) for claude-opus-4.6 and claude-opus-4.5 (percent of logged actions). Top-3 behavior shares per row are bold.}
\label{tab:behavior-frontier-claude}
\begingroup
\scriptsize
\setlength{\tabcolsep}{2pt}
\renewcommand{\arraystretch}{0.92}
\resizebox{0.90\linewidth}{!}{%
\begin{tabular}{@{}l l r r r r r r r r r@{}}
\toprule
Task & Model & Action & Spec & Plan & Read & Write & Debug & Hyg & Ext & Other\\
\midrule
\multirow[c]{2}{*}{\texttt{pug}} & claude-opus-4.6 & 528 & 6.1\% & 8.3\% & \textbf{50.8\%} & \textbf{16.1\%} & \textbf{12.9\%} & 0.2\% & 0.9\% & 4.7\% \\
 & claude-opus-4.5 & 309 & 5.8\% & 3.2\% & \textbf{27.8\%} & \textbf{35.3\%} & \textbf{20.1\%} & 0.0\% & 0.6\% & 7.1\% \\
\midrule
\multirow[c]{2}{*}{\texttt{jq}} & claude-opus-4.6 & 459 & 3.7\% & 6.5\% & \textbf{52.5\%} & \textbf{22.0\%} & \textbf{11.5\%} & 0.0\% & 2.6\% & 1.1\% \\
 & claude-opus-4.5 & 464 & 3.4\% & 3.0\% & \textbf{47.4\%} & \textbf{27.4\%} & \textbf{18.1\%} & 0.2\% & 0.0\% & 0.4\% \\
\midrule
\multirow[c]{2}{*}{\texttt{csv}} & claude-opus-4.6 & 214 & 7.9\% & 5.6\% & \textbf{30.4\%} & \textbf{27.6\%} & \textbf{26.2\%} & 0.5\% & 0.9\% & 0.9\% \\
 & claude-opus-4.5 & 149 & 4.7\% & 6.7\% & \textbf{18.1\%} & \textbf{32.9\%} & \textbf{34.2\%} & 0.0\% & 2.0\% & 1.3\% \\
\midrule
\multirow[c]{2}{*}{\texttt{ini}} & claude-opus-4.6 & 146 & 9.6\% & 6.8\% & \textbf{34.9\%} & \textbf{24.0\%} & \textbf{21.9\%} & 0.0\% & 1.4\% & 1.4\% \\
 & claude-opus-4.5 & 97 & 11.3\% & 6.2\% & \textbf{21.6\%} & \textbf{30.9\%} & \textbf{29.9\%} & 0.0\% & 0.0\% & 0.0\% \\
\midrule
\multirow[c]{2}{*}{\texttt{yaml}} & claude-opus-4.6 & 513 & 6.0\% & 4.5\% & \textbf{47.8\%} & \textbf{11.5\%} & \textbf{20.9\%} & 0.2\% & 3.3\% & 5.8\% \\
 & claude-opus-4.5 & 491 & 3.9\% & 3.1\% & \textbf{42.4\%} & \textbf{23.4\%} & \textbf{24.4\%} & 0.0\% & 0.8\% & 2.0\% \\
\midrule
\multirow[c]{2}{*}{\texttt{toml}} & claude-opus-4.6 & 3093 & 8.6\% & 5.3\% & \textbf{36.6\%} & 14.2\% & \textbf{15.3\%} & 0.2\% & \textbf{18.5\%} & 1.3\% \\
 & claude-opus-4.5 & 292 & 9.2\% & 3.8\% & \textbf{34.2\%} & \textbf{26.7\%} & \textbf{24.0\%} & 1.4\% & 0.0\% & 0.7\% \\
\midrule
\multirow[c]{2}{*}{\texttt{xml}} & claude-opus-4.6 & 542 & 6.5\% & 4.4\% & \textbf{39.5\%} & \textbf{28.0\%} & \textbf{17.3\%} & 0.0\% & 4.1\% & 0.2\% \\
 & claude-opus-4.5 & 534 & 2.6\% & 6.4\% & \textbf{32.6\%} & \textbf{26.4\%} & \textbf{31.8\%} & 0.0\% & 0.0\% & 0.2\% \\
\midrule
\multirow[c]{2}{*}{\texttt{html5}} & claude-opus-4.6 & 2179 & 5.8\% & 4.4\% & \textbf{48.5\%} & 10.3\% & \textbf{12.1\%} & 0.0\% & \textbf{14.0\%} & 4.9\% \\
 & claude-opus-4.5 & 181 & 16.0\% & 5.5\% & \textbf{16.6\%} & \textbf{36.5\%} & \textbf{23.2\%} & 0.0\% & 0.0\% & 2.2\% \\
\midrule
\multirow[c]{2}{*}{\texttt{c99}} & claude-opus-4.6 & 427 & \textbf{13.8\%} & 4.7\% & \textbf{39.1\%} & \textbf{19.9\%} & 11.0\% & 0.0\% & 0.0\% & 11.5\% \\
 & claude-opus-4.5 & 290 & 11.0\% & 5.2\% & \textbf{34.8\%} & \textbf{20.7\%} & \textbf{25.9\%} & 0.7\% & 0.0\% & 1.7\% \\
\midrule
\multirow[c]{2}{*}{\texttt{lua}} & claude-opus-4.6 & 425 & 8.0\% & 7.5\% & \textbf{38.4\%} & \textbf{21.2\%} & \textbf{20.5\%} & 0.2\% & 0.5\% & 3.8\% \\
 & claude-opus-4.5 & 465 & 2.8\% & 5.8\% & \textbf{51.4\%} & \textbf{17.4\%} & \textbf{18.9\%} & 0.0\% & 0.0\% & 3.7\% \\
\midrule
\multirow[c]{2}{*}{\texttt{ecma262}} & claude-opus-4.6 & 2384 & 6.5\% & \textbf{10.6\%} & \textbf{53.4\%} & 10.2\% & \textbf{16.9\%} & 0.0\% & 1.3\% & 1.1\% \\
 & claude-opus-4.5 & 347 & 6.1\% & 3.5\% & \textbf{42.1\%} & \textbf{30.3\%} & \textbf{13.0\%} & 0.0\% & 0.0\% & 5.2\% \\
\midrule
\multirow[c]{2}{*}{\texttt{python}} & claude-opus-4.6 & 2190 & 5.1\% & 6.0\% & \textbf{57.2\%} & \textbf{12.1\%} & \textbf{14.4\%} & 0.1\% & 1.4\% & 3.7\% \\
 & claude-opus-4.5 & 848 & 3.2\% & 2.4\% & \textbf{50.1\%} & \textbf{23.5\%} & \textbf{18.8\%} & 0.0\% & 0.9\% & 1.2\% \\
\midrule
\multirow[c]{2}{*}{\texttt{r6rs}} & claude-opus-4.6 & 3146 & 7.5\% & 5.8\% & \textbf{48.4\%} & \textbf{14.8\%} & \textbf{18.9\%} & 0.0\% & 0.4\% & 4.1\% \\
 & claude-opus-4.5 & 539 & 6.5\% & 4.6\% & \textbf{44.5\%} & \textbf{22.1\%} & \textbf{21.3\%} & 0.0\% & 0.0\% & 0.9\% \\
\midrule
\multirow[c]{2}{*}{\texttt{git\_object}} & claude-opus-4.6 & 206 & \textbf{12.6\%} & 5.3\% & \textbf{52.4\%} & \textbf{16.5\%} & 10.2\% & 0.5\% & 0.0\% & 2.4\% \\
 & claude-opus-4.5 & 261 & 10.0\% & 3.4\% & \textbf{33.7\%} & \textbf{24.1\%} & \textbf{24.1\%} & 0.0\% & 0.0\% & 4.6\% \\
\midrule
\multirow[c]{2}{*}{\texttt{protobuf}} & claude-opus-4.6 & 97 & 14.4\% & 8.2\% & \textbf{27.8\%} & \textbf{32.0\%} & \textbf{17.5\%} & 0.0\% & 0.0\% & 0.0\% \\
 & claude-opus-4.5 & 94 & \textbf{26.6\%} & 16.0\% & \textbf{20.2\%} & \textbf{26.6\%} & 10.6\% & 0.0\% & 0.0\% & 0.0\% \\
\midrule
\multirow[c]{2}{*}{\texttt{zip}} & claude-opus-4.6 & 1990 & 3.6\% & 6.5\% & \textbf{57.3\%} & 7.1\% & \textbf{10.2\%} & 0.0\% & 4.9\% & \textbf{10.5\%} \\
 & claude-opus-4.5 & 595 & 4.9\% & 1.2\% & \textbf{44.2\%} & \textbf{21.3\%} & \textbf{19.7\%} & 0.2\% & 0.0\% & 8.6\% \\
\midrule
\multirow[c]{2}{*}{\texttt{capnp}} & claude-opus-4.6 & 195 & \textbf{16.4\%} & 5.6\% & \textbf{39.0\%} & \textbf{26.2\%} & 11.8\% & 0.5\% & 0.0\% & 0.5\% \\
 & claude-opus-4.5 & 87 & \textbf{21.8\%} & 8.0\% & \textbf{36.8\%} & \textbf{20.7\%} & 11.5\% & 1.1\% & 0.0\% & 0.0\% \\
\midrule
\multirow[c]{2}{*}{\texttt{wasm}} & claude-opus-4.6 & 211 & \textbf{22.7\%} & 5.2\% & \textbf{29.4\%} & \textbf{19.9\%} & 17.1\% & 0.0\% & 0.5\% & 5.2\% \\
 & claude-opus-4.5 & 450 & 6.7\% & 2.2\% & \textbf{32.2\%} & \textbf{24.0\%} & \textbf{34.0\%} & 0.0\% & 0.0\% & 0.9\% \\
\midrule
\multirow[c]{2}{*}{\texttt{uri}} & claude-opus-4.6 & 101 & \textbf{19.8\%} & 5.9\% & \textbf{27.7\%} & \textbf{24.8\%} & 18.8\% & 1.0\% & 0.0\% & 2.0\% \\
 & claude-opus-4.5 & 124 & \textbf{22.6\%} & 6.5\% & \textbf{22.6\%} & \textbf{25.0\%} & 21.8\% & 0.8\% & 0.0\% & 0.8\% \\
\midrule
\multirow[c]{2}{*}{\texttt{hpack}} & claude-opus-4.6 & 136 & 11.8\% & 8.8\% & \textbf{27.2\%} & \textbf{24.3\%} & 8.8\% & 0.0\% & 5.1\% & \textbf{14.0\%} \\
 & claude-opus-4.5 & 143 & \textbf{22.4\%} & 6.3\% & \textbf{25.9\%} & \textbf{25.2\%} & 19.6\% & 0.0\% & 0.0\% & 0.7\% \\
\midrule
\multirow[c]{2}{*}{\texttt{url}} & claude-opus-4.6 & 432 & \textbf{22.7\%} & 4.4\% & \textbf{22.0\%} & \textbf{16.0\%} & 14.6\% & 0.2\% & 12.7\% & 7.4\% \\
 & claude-opus-4.5 & 290 & 11.0\% & 5.9\% & \textbf{29.7\%} & \textbf{31.4\%} & \textbf{20.0\%} & 0.0\% & 0.0\% & 2.1\% \\
\midrule
\multirow[c]{2}{*}{\texttt{cdcl}} & claude-opus-4.6 & 772 & 6.0\% & 7.5\% & \textbf{44.3\%} & \textbf{15.9\%} & \textbf{22.4\%} & 0.1\% & 2.3\% & 1.4\% \\
 & claude-opus-4.5 & 334 & 2.7\% & 8.4\% & \textbf{32.6\%} & \textbf{27.2\%} & \textbf{28.7\%} & 0.0\% & 0.0\% & 0.3\% \\
\bottomrule
\end{tabular}%
}
\endgroup
\end{table}

\end{document}